\documentclass[prd,aps,letterpaper,floatfix,superscriptaddress,preprintnumbers,11pt,nofootinbib]{revtex4-1}
\usepackage{epsfig,epsf}
\usepackage{booktabs}
\usepackage{mathrsfs}
\usepackage{graphicx}
\usepackage{subcaption}
\usepackage{caption}
\usepackage{latexsym}
\usepackage{amsmath}
\usepackage{amssymb}
\usepackage{textcomp}
\usepackage{amsbsy}
\usepackage{tabularx}
\usepackage{slashed}
\usepackage{array}
\usepackage[dvipsnames]{xcolor}
\usepackage{bm}
\usepackage{amsmath,amsfonts,amssymb}
\usepackage{bbold}
\usepackage{physics}
\usepackage{hyperref}
\DeclareGraphicsExtensions{.png,.eps,.pdf}

\newcommand{\arXivold}[2]{\href{http://arxiv.org/pdf/#1}{{\tt #2/#1}}}

\usepackage{float}

\usepackage{color}
\usepackage[normalem]{ulem}

\definecolor{darkgreen}{cmyk}{1,0,1,0.4}

\long\def\/*#1*/{}

\begin{document}
\title{Flavor-violating Di-Higgs Couplings}
\author{Fayez Abu-Ajamieh}
\email{fayezajamieh@iisc.ac.in}
\affiliation{Centre for High Energy Physics, Indian Institute of Science, Bangalore 560012, India}
\author{Marco Frasca}
\email{marcofrasca@mclink.it}
\affiliation{Rome, Italy}
\author{Sudhir K. Vempati}
\email{vempati@iisc.ac.in}
\affiliation{Centre for High Energy Physics, Indian Institute of Science, Bangalore 560012, India}
\begin{abstract}
Di-Higgs couplings to fermions of the form $h^{2}\overline{f}f$ are absent in the Standard Model, however, they are present in several physics Beyond Standard Model (BSM) extensions, including those with vector-like fermions.  In Effective Field  Theories (EFTs), such as the Standard Model Effective Field Theory (SMEFT) and the Higgs Effective Field Theory (HEFT), these couplings appear at dimension 6 and can in general, be flavor-violating (FV). In the present work, we employ a bottom-up  approach to investigate the FV in the lepton and quarks sectors through the di-Higgs effective couplings. We assume that all FV arises from this type of couplings and assume that the Yukawa couplings $Y_{ij}$ are given by their SM values, i.e., $Y_{ij} = \sqrt{2}m_{i}\delta_{ij}/v$. In the lepton sector, we set upper limits on the Wilson coefficients $C_{ll'}$ from $l \rightarrow 3l'$ decays, $l \rightarrow l\gamma$ decays, muonium oscillations, the $(g-2)_{\mu}$ anomaly, LEP searches, muon conversion in nuclei, FV Higgs decays, and $Z$ decays. We also make projections on some of these coefficients from Belle II, the Mu2e experiment and the LHC's High Luminosity (HL) run. In the quark sector, we set upper limits on the Wilson coefficients $C_{qq'}$ from meson oscillations and from $B$-physics searches. A key takeaway from this study is that current and future experiments should set out to measure the effective di-Higgs couplings $C_{ff'}$, whether these couplings are FV or flavor-conserving. We also present a matching between our formalism and the SMEFT operators and show the bounds in both bases. 
\end{abstract}
\maketitle

\section{Introduction}

Flavor physics provides an essential probe for the Standard Model (SM) and for new physics BSM. In the SM, flavor violation (FV) arises entirely through the fermionic couplings to the Higgs bosons, i.e., through the Yukawa matrices. These Yukawa matrices encode FV in the CKM matrix in the hadronic sector, and in the UPMNS matrix in the leptonic sector. In physics BSM, any new source of flavor violation is severely constrained. FV processes are well measured in $\Delta F =1$ and $\Delta F=2$ transitions. Some of the most robust constraints are obtained from $K^0-\overline{K}^0$ system in the quark sector, and from $\mu \to e + \gamma$ in the leptonic sector. Other processes which are not flavor violating ($\Delta F=0$) but still play an essential role in constraining new physics, are the magnetic and electric dipole moments of leptons, nucleons, atoms, and molecules. To avoid strong constraints on new physics from flavor physics, typically it is assumed to follow the paradigm of Minimal Flavor Violation (MFV)\cite{DAmbrosio:2002vsn}.  

An interesting scenario would arise when non-minimal FV is induced through the effective Higgs couplings to fermions. There are many new physics scenarios  where non-minimal FV can arise through the Higgs couplings, such as the multi-Higgs models, the Randall-Sundrum models and so on. The case of FV couplings with a single Higgs has been studied in Ref. \cite{Dery:2013rta,Harnik:2012pb}. FV can be understood in terms of deviations of the SM Yukawa couplings from their SM values in the generation space. A complete global analysis of flavor observables was performed and the limits on the FV Yukawa couplings were derived. This work is similar in theme to the analysis conducted in \cite{Dery:2013rta, Harnik:2012pb}, and extends it to the case of FV through the di-Higgs couplings to fermions.

Di-Higgs-fermion-fermion couplings are absent in the SM; however, they can be generated in a way similar to the single Higgs couplings in many new physics scenarios. A simple example of this are extensions of the SM with extra vector-like fermions. In the limit of heavy vector-like fermions, integrating them out would lead to operators with di-Higgs couplings to the SM fermions\footnote{These are not the only set of operators after integrating the heavy fermions. But we focus on these operators for the present discussion.}. These operators can be mapped to EFT frameworks, such as the HEFT and the SMEFT, at the level of dimension six operator (see for example, \cite{deBlas:2017xtg,delAguila:2000aa,Chen:2017hak,Batell:2012ca} and the references therein). The study of FV in EFTs has been performed in many works in the literature, see for example \cite{Silvestrini:2018dos,Descotes-Genon:2018foz,Aebischer:2018iyb,Greljo:2023adz,Greljo:2022cah,Bruggisser:2021duo,Aoude:2020dwv,Hurth:2019ula, Calibbi:2021pyh, Ali:2023kua}. To the best of our knowledge, non-minimal FV di-Higgs couplings have never been studied previously in the literature, as in most cases, these $h^{2}\overline{f}f$ operators are either avoided entirely or assumed to be proportional the Yukawa couplings by imposing (minimal) flavor symmetries \cite{Aebischer:2020lsx,Faroughy:2020ina}.  

Non-minimal di-Higgs couplings are interesting, as they have unique signatures, and can be probed by future colliders, especially the muon collider. 
A non-minimal di-Higgs coupling could even explain the discrepancy of the muon $g-2$ anomaly \cite{Abu-Ajamieh:2022nmt}.
In studying these couplings in the present work, we find it suitable to follow the framework proposed in \cite{Chang:2019vez, Abu-Ajamieh:2020yqi, Abu-Ajamieh:2021egq, Abu-Ajamieh:2022ppp, Abu-Ajamieh:2021vnh}. We call this framework the Weak Scale Deviations framework (WSD). This formalism is model-independent and bottom-up, as it considers all possible deviations from the SM Lagrangian. The FV di-Higgs couplings appear naturally in the expansion of the Higgs operator in this formalism, along with deviations in the Yukawa couplings. While one could choose to work within either the SMEFT or the HEFT, we find the WSD framework to be more convenient and advantageous, as it has fewer assumptions compared to either the SMEFT or the HEFT and is more closely-linked to experiment as we show later on. Nonetheless, we shall present the mapping of the WSD to the SMEFT and present the SMEFT cutoff scale that corresponds to the upper limits on the FV di-Higgs Wilson coefficients for convenience. 

Focusing on the di-Higgs couplings, we provide a complete analysis of the flavor physics constraints for both the quark and the lepton sectors. Our analysis follows similar lines as the analysis performed in \cite{Harnik:2012pb} for FV Higgs Yukawa couplings. The results for the di-Higgs couplings are presented in terms of the bounds on the  Wilson coefficient of the $h^{2}\overline{f}f$ operators and also on the corresponding UV scale in the SMEFT. The bounds on the SMEFT operators are competitive and are similar to those on new physics. For example, assuming the Wilson coefficients to be $\mathcal{O}(1)$ in the SMEFT, the bounds on the UV scale $\Lambda$ range from $\sim 1 - 10$ TeV in the leptonic sector and can exceed $100$ TeV in the  $K^0-\overline{K^0}$ oscillations in the quark sector. 

This paper is organized as follows: In Section {\rm II}, we briefly review the WSD formalism we utilize in this paper. In Section {\rm III}, we present our complete analysis on the FV through the di-Higgs couplings in the leptonic sector, whereas in Section {\rm IV}, we do the same analysis in the quark sector. In Section {\rm V} and show how this formalism can be mapped to the SMEFT framework, and in particular derive the UV scale that corresponds to the upper limit on the FV Wilson coefficient. Finally, we present our conclusions in Section {\rm VI}. We relegated much of the calculational details to the appendices \textbf{A - D}.

\section{Framework}
We begin by introducing our FV framework, which is essentially based on the phenomenological bottom-up WSD approach introduced in \cite{Chang:2019vez, Abu-Ajamieh:2020yqi, Abu-Ajamieh:2021egq,Abu-Ajamieh:2022ppp, Abu-Ajamieh:2021vnh}, generalized to the case of FV couplings and Wilson coefficients. In this framework, we avoid power expansion in writing down higher-dimensional operators, as the case in the SMEFT. Instead, we parameterize New Physics (NP) as deviations from the SM predictions without making any references to any UV scale. Therefore, we write the most general FV effective Lagrangian of the Yukawa interaction as follows 

\begin{equation}\label{eq:BSM_Lag1}
\mathcal{L}_{\text{eff}} =  -\frac{v}{\sqrt{2}}\Big( \overline{L}_{l}^{i}\Tilde{\hat{H}} l_{R}^{j} + \text{h.c.}\Big)\Bigg[ Y^{l}_{ij} \frac{X}{v} + C^{l}_{ij}\frac{X^{2}}{2!v^{2}}+ \dots \Bigg] -\frac{v}{\sqrt{2}}\Big( \overline{Q}_{l}^{i}\Tilde{\hat{H}} q_{R}^{j} + \text{h.c.}\Big)\Bigg[ Y^{q}_{ij} \frac{X}{v} + C^{q}_{ij}\frac{X^{2}}{2!v^{2}}+ \dots \Bigg],
\end{equation}
where $Y^{l}_{ij}$ and $Y^{q}_{ij}$ are the Yukawa coupling matrices for the leptons and the quarks, respectively, whereas $C^{l}_{ij}$ and $C^{q}_{ij}$ are matrices containing FV Wilson coefficients that do not have SM counterparts. Also notice that in the SM we have $Y^{l,q}_{ij} = \delta_{ij}\sqrt{2} m_{i}/v$, and $C^{l,q}_{ij}=0$. The field $X$ is defined in terms of the Higgs doublet $H$ as
\begin{equation}\label{eq:Xfield}
X = \sqrt{2H^{\dagger}H} - v,
\end{equation}
whereas we define the projector $\Tilde{\hat{H}} = \epsilon \hat{H}^{\ast}$, with
\begin{equation}\label{eq:Projector}
\epsilon = \begin{pmatrix} 
            0 && 1 \\ 
            -1 && 0
            \end{pmatrix}, \hspace{10mm}
            \hat{H} = \frac{H}{\sqrt{H^{\dagger}H}} = \begin{pmatrix} 0 \\ 
                                                                    1 
                                                    \end{pmatrix} + O(\vec{G}),
\end{equation}
where $\vec{G}$ are the Goldstone bosons. Notice that $X$ has the same quantum numbers as the Higgs field, and in the unitary gauge we have $X \rightarrow h$. Before we proceed, a few of remarks are in order. 
\begin{itemize}
\item Notice that in Eq. \ref{eq:BSM_Lag1}, we are dividing the field $X$ by appropriate powers of $v$ in order to keep Wilson coefficients dimensionless, i.e., $v$ should not be interpreted as an expansion scale as the case in the HEFT \cite{Grinstein:2007iv}, and the Wilson coefficients could in principle assume any value allowed by unitarity and experiment, 
\item We are assuming that $v$ is the minimum of Higgs potential including all higher-order corrections. Therefore, $v=246$ GeV. In addition, the value Higgs mass remains equal to the measured one, i.e., $125$ GeV,
\item Although Eq. (\ref{eq:BSM_Lag1}) appears to be similar to the HEFT, we should keep in mind that secretly we are using the Higgs doublet in our expansion, and one can easily demonstrate that the effective Lagrangian in Eq. (\ref{eq:BSM_Lag1}) can be mapped to either the SMEFT or the HEFT, depending on the chosen expansion, i.e. Eq. (\ref{eq:BSM_Lag1}) can be mapped to SMEFT when $X \rightarrow H$, and can be mapped to HEFT when $X \rightarrow h$, as the case when the unitary gauge is chosen, AND when $v$ is interpreted as a true expansion scale. In either the SMEFT or the HEFT frameworks, the deviations and Wilson coefficients in eqs. (\ref{eq:BSM_Lag1}) can receive corrections from a tower of higher-order operators, which might be different depending on the order at which we truncate the expansion. We will present the matching to the SMEFT in Section \ref{sec:SMEFT} below and show the corresponding scale of NP. The interested reader is instructed to refer to \cite{Abu-Ajamieh:2020yqi, Abu-Ajamieh:2021egq,Abu-Ajamieh:2022ppp, Abu-Ajamieh:2021vnh} for more details on mapping the operators into the SMEFT and the HEFT.

\item There are two advantages to this construction: First, there are fewer assumptions in this framework compared to either the SMEFT or the HEFT. Namely, we are only assuming that there are no light degrees of freedom below the energy scale at which the EFT breaks down, and that the deviations and Wilson coefficients are compatible with experimental measurements. The second benefit lies in the fact that parameterizing NP this way is more transparent phenomenologically, and more closely linked to experiment, as these deviations and Wilson coefficients are what is measured experimentally as opposed to any expansion scale. 
\end{itemize}

\begin{figure}[!t]
\centering
\includegraphics[width = 0.6\textwidth]{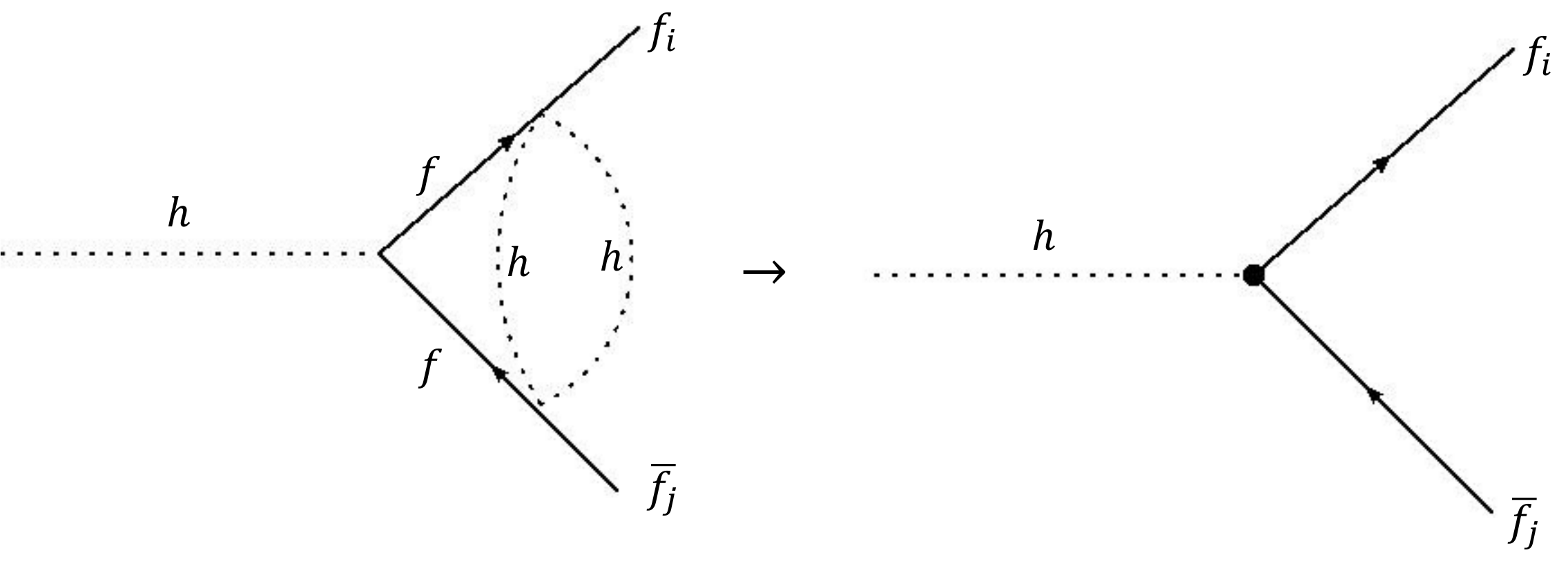}
\caption{\small Induced FV Yukawa couplings $Y^{\text{ind}}_{ij}$ through non-zero $C_{ij}$.}
\label{fig0}
\end{figure}
 It is commonly assumed in the literature that $Y_{ij}$ are the main source of FV, and studies that investigate limits on  $Y_{ij}$ abound (see for instance \cite{Dery:2013rta, Harnik:2012pb,Zhang:2021nzv,Vicente:2019ykr,Soreq:2016rae,Buschmann:2016uzg}. In this paper however, we are more interested in the case where the effective couplings $C_{ij}$ are the main source of FV. Therefore, we assume
 \begin{equation}\label{eq:SM_Yukawas}
 Y_{ij} \simeq Y^{\text{SM}}_{ij} = \frac{\sqrt{2}m_{i}}{v}\delta_{ij},
 \end{equation}
for both the quarks and the leptons. We call FV through the couplings $h^{2}\overline{f}f$ the next-to-minimum FV through di-Higgs effective couplings. The reason why it is not possible to make $Y_{ij} = Y^{\text{SM}}_{ij}$ exactly, is that it is not possible to simultaneously diagonalize both $Y_{ij}$ and $C_{ij}$, as non-zero $C_{ij}$ will induce corrections to $Y_{ij}$ at 2-loops as shown in Fig. \ref{fig0}. Let's call this part of the Yukawas $Y^{\text{ind}}_{ij}$ to distinguish it from any corrections arising from any other source. We can estimate the size of $Y^{\text{ind}}_{ij}$ as follows
\begin{equation}\label{eq:estimated_Yij}
    Y^{\text{ind}}_{ij} \sim \Big( \frac{1}{16\pi^{2}} \Big)^{2} \times Y_{ff} C_{ff_{i}}C_{ff_{j}},
\end{equation}
which for $C_{ff_{i}}, C_{ff_{j}} \sim O(1)$ implies that $Y^{\text{ind}}_{ij} \sim O(10^{-5})$ at best, i.e. the FV contributions from $Y^{\text{ind}}_{ij}$ are always suppressed compared to those arising from $C_{ij}$ and are thus negligible. We will not concern ourselves with these corrections in the remainder of this paper.

In the unitary gauge, the FV part of Eq. \ref{eq:BSM_Lag1} reads
\begin{equation}\label{FV_Cij}
\mathcal{L}_{\text{FV}} \supset -\frac{C^{l}_{ij}}{2\sqrt{2}v}\Big(\overline{l}^{i}_{L}l^{j}_{R} +\text{h.c.} \Big)h^{2} -\frac{C^{q}_{ij}}{2\sqrt{2}v}\Big(\overline{q}^{i}_{L}q^{j}_{R} +\text{h.c.} \Big)h^{2}.
\end{equation}

In general, the matrices $C^{l,q}_{ij}$ could be complex and needn't be symmetric. However, in this paper, we will simplify by assuming that they are both real and symmetric, i.e., $\text{Im}(C^{l,q}_{ij}) = 0$ and $C^{l,q}_{ij} = C^{l,q}_{ji}$.

\section{The Lepton Sector}\label{sec:lepton}
We focus first on FV in the lepton sector. Explicitly, the lepton part of Eq. (\ref{FV_Cij}) reads
\begin{equation}\label{eq:lepton_FV}
\mathcal{L}^{l}_{\text{FV}} \supset -\frac{1}{2\sqrt{2}v}\Big[C_{ee}\overline{e}e + C_{\mu\mu}\overline{\mu}\mu + C_{\tau \tau}\overline{\tau}{\tau} + C_{\mu e}(\overline{e}\mu + \overline{\mu}e) + C_{\tau\mu}(\overline{\mu}\tau + \overline{\tau}\mu) + C_{\tau e}(\overline{e}\tau + \overline{\tau}e)\Big]h^{2}.
\end{equation}
Notice that $C_{ll}$ are not FV, however, they will enter into the calculation and bounds along with the FV couplings $C_{ll'}$. The bounds on are summarized in Table \ref{table1} and shown in Figures \ref{fig8} and \ref{fig9}. Below, we discuss these bounds in more detail.

\begin{table}[!ht]
\centering
\vspace{1 mm}
\tabcolsep7pt\begin{tabular}{c c c c}
\hline
\hline
 \textbf{Channel} & \textbf{Couplings} &  \textbf{Bounds ($\Lambda$ TeV)} & \textbf{Projections ($\Lambda$ TeV)} \\
\hline
\hline
$\tau \rightarrow 3\mu$ & $|C_{\tau\mu}||C_{\mu\mu}|$ & $< 2.54 \times 10^{-2}$ $(> 1.07)$ & $< 3.92 \times 10^{-3}$ $(> 1.7)$ \\
$\mu \rightarrow 3e$ & $|C_{\mu e}||C_{ee}|$ & $< 4.41 \times 10^{-5}$ $(> 5.23)$ & $< 4.41 \times 10^{-7}$ $(> 16.53)$\\
$\tau \rightarrow 3e$ & $|C_{\tau e}||C_{ee}|$ & $< 2.88 \times 10^{-2}$ $(> 1.03)$ & $< 3.92\times 10^{-3}$ $(> 1.7)$\\
$\tau^{-} \rightarrow e^{+}\mu^{-}\mu^{-}$ & $|C_{\tau \mu}||C_{\mu e}|$ & $< 2.29 \times 10^{-2}$ $(> 1.1)$ & $< 2.83\times 10^{-3}$ $(>1.85)$ \\
$\tau^{-} \rightarrow \mu^{+}e^{-}e^{-}$ & $|C_{\tau e}||C_{\mu e}|$ & $< 2.15 \times 10^{-2}$ $(>1.11)$ & $< 2.66\times 10^{-3}$ $(>1.88)$\\
$\tau^{-} \rightarrow \mu^{+}\mu^{-}e^{-}$ & $|C_{\tau \mu}||C_{\mu e}|$, $|C_{\tau e}||C_{\mu \mu}|$ & $< 2.88 \times 10^{-2}$ $(>1.03)$ & $< 3.72\times 10^{-3}$ $(>1.73)$\\
$\tau^{-} \rightarrow \mu^{-}e^{+}e^{-}$ & $|C_{\tau \mu}||C_{ee}|$, $|C_{\tau e}||C_{\mu e}|$ & $< 2.35 \times 10^{-2}$ $(>1.09)$ & $< 2.99\times 10^{-3}$ $(>1.82)$ \\
\hline
$\mu \rightarrow e \gamma$ ($\tau$ in loop)& $|C_{\tau\mu}||C_{\tau e}|$ & $ < 7.83 \times 10^{-5} $ $(> 4.53)$ & $< 2.7 \times 10^{-5}$ $(> 5.91)$\\
$\mu \rightarrow e \gamma$ ($\mu$ in loop) & $|C_{\mu\mu}||C_{\mu e}|$ & $ < 4.4 \times 10^{-4} $ $(> 2.94)$ & $< 1.52 \times 10^{-4}$ $(> 3.84)$\\
$\mu \rightarrow e \gamma$ ($e$ in loop)& $|C_{\mu e}||C_{e e}|$ & $ < 8.28 \times 10^{-4} $ $(> 2.51)$ & $< 2.86 \times 10^{-4}$ $(> 3.28)$\\
$\tau \rightarrow \mu \gamma$ ($\tau$ in loop) & $|C_{\tau\tau}||C_{\tau\mu}|$ & $ < 0.66 $ $(> 0.47)$& $< 9.92 \times 10^{-2}$ $(> 0.76)$\\
$\tau \rightarrow \mu \gamma$ ($\mu$ in loop)& $|C_{\tau\mu}||C_{\mu\mu}|$ & $ < 1.12 $ $(> 0.41)$ & $< 0.17$ $(> 0.66)$\\
$\tau \rightarrow \mu \gamma$ ($e$ in loop)& $|C_{\tau e}||C_{\mu e}|$ & $ < 0.64 $ $(> 0.48)$ & $< 9.66 \times 10^{-2}$ $(> 0.76)$\\
$\tau \rightarrow e \gamma$ ($\tau$ in loop)& $|C_{\tau\tau}||C_{\tau e}|$ & $ < 0.57 $ $(> 0.49)$ & $< 0.22$ $(>0.62)$\\
$\tau \rightarrow e \gamma$ ($\mu$ in loop) & $|C_{\tau\mu}||C_{\mu e}|$ & $ < 0.97 $ $(> 0.43)$ & $< 0.38$ $(> 0.54)$\\
$\tau \rightarrow e \gamma$ ($e$ in loop) & $|C_{\tau e}||C_{e e}|$ & $ < 0.55 $ $(> 0.49)$ & $< 0.22 $ $(> 0.62)$\\
\hline
$M-\overline{M}$ oscillations & $|C_{\mu e}|$ & $ < 0.39$ $(> 0.68)$ & -\\
\hline
$(g-2)_{\mu}$ & $|C_{\tau\mu}|$ & $0.26 \pm 0.03$ $(> 0.84)$ & -\\
$(g-2)_{\mu}$ & $|C_{\mu\mu}|$ & $0.79 \pm 0.1$ $(> 0.48)$ & -\\
$(g-2)_{\mu}$ & $|C_{\mu e}|$ & $6.34 \pm 0.8$ $(> 0.17)$ & -\\
\hline
LEP & $|C_{\tau e}|$ & $< 9.52$ $(> 0.14)$ & - \\
LEP & $|C_{\mu e}|$ & $< 9.0$ $(> 0.14)$ & - \\
LEP & $|C_{e e}|$ & $< 13.25$ $(> 0.12)$ & - \\
\hline
$\mu \rightarrow e$ conversion in nuclei & $|C_{\mu e}|$ & $< 0.34$ $(> 0.73)$ & $<4.56 \times 10^{-3}$ $(> 6.31)$\\
\hline
$h \rightarrow \tau \mu$ & $|C_{\tau\mu}|$ & $<0.67$ $(> 0.52)$ & $<0.23$ $(> 0.89)$\\
$h \rightarrow \tau e$ & $|C_{\tau e}|$ & $<1.04$ $(> 0.42)$ & $<0.23$ $(> 0.89)$\\
$h \rightarrow  \mu e$ & $|C_{\mu e}|$ & $<0.25$ $(> 0.85)$ & $<7.3 \times 10^{-2}$ $(> 1.58)$ \\
$h \rightarrow e e$ & $|C_{ee}|$ & $<0.58$ $(> 0.56)$ & - \\
\hline
$Z \rightarrow \tau^{+}\tau^{-}$ & $|C_{\tau\tau}|$, $|C_{\tau\mu}|$, $|C_{\tau e}|$ & $< 7.9$ $(> 0.15)$ & -\\
$Z \rightarrow \mu^{+}\mu^{-}$ & $|C_{\tau\mu}|$, $|C_{\mu\mu}|$, $|C_{\mu e}|$ & $< 7.04$ $(> 0.16)$ & -\\
$Z \rightarrow e^{+}e^{-}$ & $|C_{\tau e}|$, $|C_{\mu e}|$, $|C_{e e}|$ & $< 5.62$ $(> 0.18)$ & -\\
\hline
$Z \rightarrow \tau^{\pm}\mu^{\mp}$ & $|C_{\tau\mu}|$ & $< 0.11$ $(> 1.28)$ & -\\
$Z \rightarrow \tau^{\pm}e^{\mp}$ & $|C_{\tau e}|$ & $< 9.65 \times 10^{-2}$ $(> 1.37)$ & -\\
$Z \rightarrow \mu^{\pm}e^{\mp}$ & $|C_{\mu e}|$ & $< 1.59 \times 10^{-3}$ $(> 10.69)$ & -\\
\hline
\hline
\end{tabular}
    \caption{\label{table1} \small $90\%$ CL bounds and projections on the leptonic next-to-minimal FV di-Higgs couplings and the corresponding lower limit on the scale of NP $\Lambda$ from matching to the SMEFT.}
\label{table1}
\end{table}

\subsection{Bounds from $l \rightarrow l_{1}l_{2}l_{3}$ decays}\label{sec:lto3l}
\begin{figure}[!t]
\centering
\includegraphics[width = 0.3\textwidth]{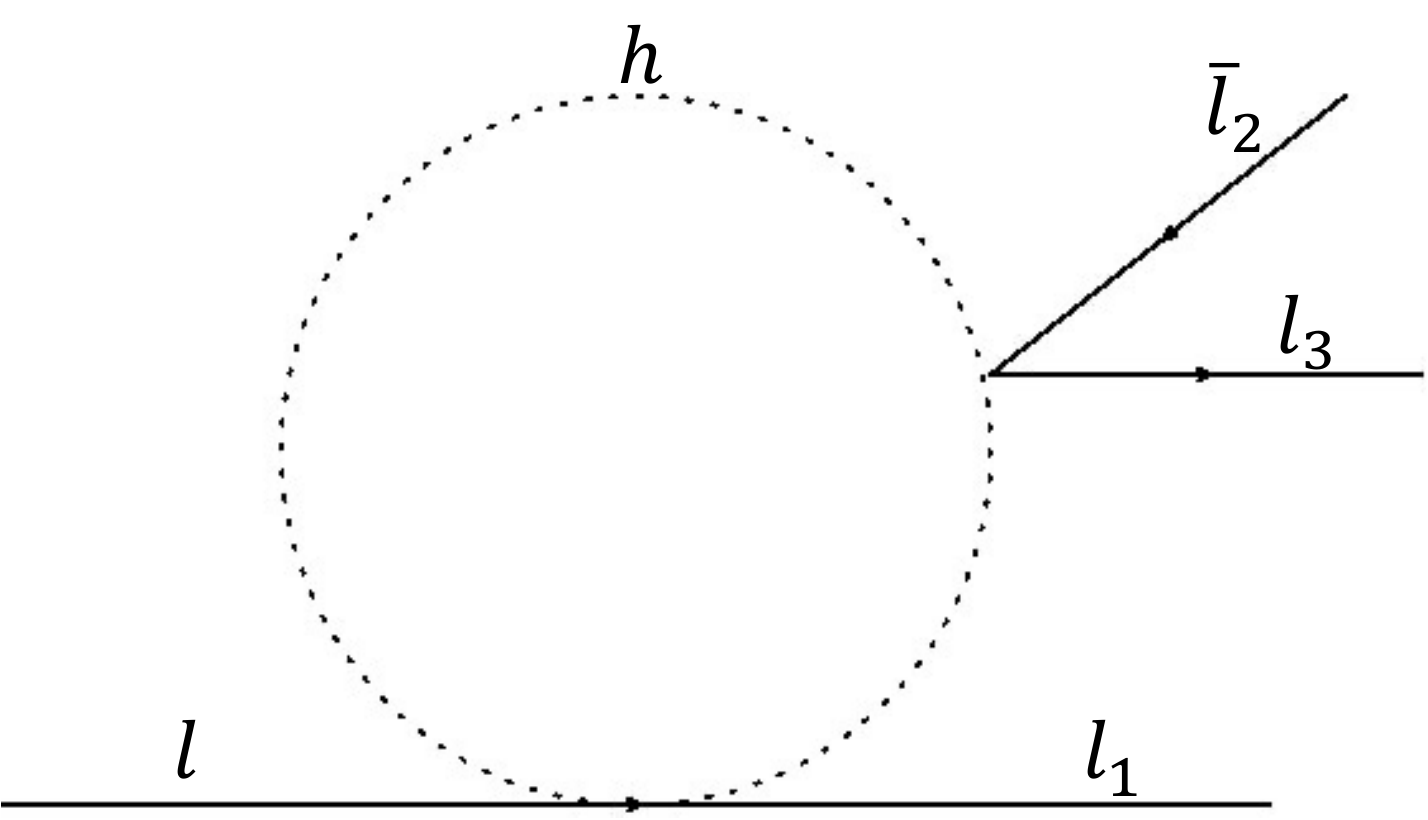}
\caption{\small The $l \rightarrow l_{1}l_{2}l_{3}$ decay through the di-Higgs effective couplings $C_{ij}$.}
\label{fig1}
\end{figure}

The $l \rightarrow l_{1}l_{2}l_{3}$ decay through the di-Higgs couplings proceeds at one loop as in Figure \ref{fig1}. Here, the $h^{2}ll'$ vertices should be viewed as effective interactions of some heavy degree(s) of freedom that has been integrated out. In the limit $M_{h} \gg m_{l}$, the decay width can be approximated as
\begin{equation}\label{lto3lDecay}
   \Gamma(l \rightarrow l_{1}l_{2}l_{3}) \simeq \frac{m_{l}^{5}}{v^{4}}\Bigg[\frac{C_{l l_{1}}C_{l_{2}l_{3}}}{512\pi^{3}\sqrt{6\pi}} \log{\Big( \frac{M_{h}^{2}}{m_{l}^{2}}\Big)} \Bigg]^{2}.
\end{equation}

The detailed calculation is given in Appendix \ref{appendix1}. Before we proceed with extracting the bounds, we should note that the 2-loop diagram (similar to the bottom diagram in Figure \ref{fig2}, with the photon decaying to $l_{2}l_{3}$) is suppressed relative to the 1-loop diagram and can be neglected. 

The relevant processes are $\tau \rightarrow 3\mu$, $\mu \rightarrow 3e$, $\tau \rightarrow \mu\mu e$, $\tau \rightarrow \mu ee$ and $\tau \rightarrow 3e$. The latest bounds on the branching rations of these processes can be found in \cite{ParticleDataGroup:2018ovx}, and all of them are given $@$ $90\%$ C.L., which we stick to throughout this paper. 
For the first process, the experimental bound is $\text{Br}(\tau \rightarrow 3\mu) < 2.1 \times 10^{-8}$, which translates into the bound $|C_{\tau\mu}||C_{\mu\mu}| < 2.54 \times 10^{-2}$. Notice that the FV coupling $|C_{\tau\mu}|$ cannot be isolated from the non-FV one $|C_{\mu\mu}|$. This is a common feature of these types of couplings. The second experimental limit is given by $\text{Br}(\mu \rightarrow 3e) < 1 \times 10^{-12}$, which translates to the $|C_{\mu e}||C_{ee}| < 4.41 \times 10^{-5}$. The limit on the third process is $\text{Br}(\tau^{-}\rightarrow \mu^{-}\mu^{-}e^{+}) < 1.7 \times 10^{-8}$, which translates into $|C_{\tau\mu}||C_{\mu e}| < 2.29 \times 10^{-2}$. The limit on the fourth decay is $\text{Br}(\tau \rightarrow 3e) < 2.7 \times 10^{-8}$, yielding the bound $|C_{\tau e}||C_{ee}| < 2.88 \times 10^{-2}$. The bounds on the fifth process read $\text{Br}(\tau^{-} \rightarrow \mu^{+}e^{-}e^{-}) < 1.5 \times 10^{-8}$ and translate into the limit $|C_{\tau e}||C_{\mu e}| < 2.15 \times 10^{-2}$. 

The last 2 decays are more subtle as they involve two Feynman diagrams instead of one. The decay width is obtained by summing two matrix element which have different FV couplings. For the decay $\tau^{-} \rightarrow \mu^{+}\mu^{-}e^{-}$, in the first diagram, we have $l = \tau^{-}$, $l_{1} = e^{-}$, $l_{2} = \mu^{+}$, $l_{3} =\mu^{-}$, whereas in the second we have $l = \tau^{-}$, $l_{1} = \mu^{-}$, $l_{2} = \mu^{+}$, $l_{3} =e^{-}$. The experimental limit is $\text{Br}(\tau^{-} \rightarrow \mu^{+}\mu^{-}e^{-}) < 2.7 \times 10^{-8}$, which translates into the bound $[2C_{\tau\mu}^{2}C_{\mu e}^{2}+ 2C_{\mu\mu}^{2}C_{\tau e}^{2} - C_{\tau\mu}C_{\tau e}C_{\mu\mu}C_{\mu e}]^{1/2} < 4.07 \times 10^{-2}$. Upper bounds can be obtained by setting $C_{\tau \mu} = C_{\mu e} = 0$ ($C_{\tau e} = C_{\mu \mu} = 0$) in the first (second) diagrams, which yields the bounds $|C_{\tau \mu}| |C_{\mu e}|, |C_{\tau e}| |C_{\mu \mu}|< 2.88 \times 10^{-2}$. In the final process $\tau^{-} \rightarrow \mu^{-}e^{+}e^{-}$, the two Feynman diagrams are given by $l = \tau^{-}$, $l_{1} = e^{-}$, $l_{2} = e^{+}$, $l_{3} = \mu^{-}$ in the first diagram, and $l = \tau^{-}$, $l_{1} = \mu^{-}$, $l_{2} = e^{+}$, $l_{3} = e^{-}$. The experimental bound for this process is $\text{Br}(\tau^{-} \rightarrow \mu^{-}e^{+}e^{-}) < 1.8 \times 10^{-8}$, which translates into the mixed bound $[2C_{\tau \mu}^{2}C_{ee}^{2} + 2C_{\tau e}^{2} C_{\mu e}^{2}- C_{\tau \mu}C_{\tau e}C_{\mu e}C_{ee}]^{1/2} < 3.33 \times 10^{-3}$, from which the upper bounds $|C_{\tau\mu}||C_{ee}|, |C_{\tau e}||C_{\mu e}|< 2.35 \times 10^{-2}$ are obtained.

Better bounds can be obtained from future experiments. In particular, the Belle II experiment \cite{Aushev:2010bq, Belle-II:2022cgf} is expected to collect $50 \hspace{1mm} \text{ab}^{-1}$ over the next decade, and the bounds on the branching rations of the above processes are projected to be $\sim O(10^{-10})$ (see also \cite{Calibbi:2017uvl, Banerjee:2022vdd}\footnote{The projections provided in these two references are slightly different. For our projected limits, we use the stronger of the two.}). This leads to bounds that are 1-2 orders of magnitude stronger that what is currently available. For instance, the projected bound from Belle II for $\text{Br}(\tau \rightarrow 3\mu)$ is $5\times 10^{-10}$. This yields the projected bound $|C_{\tau\mu}||C_{\mu\mu}|<3.92 \times 10^{-3}$. The rest of the projections are summarized in Table \ref{table1}.

\subsection{Bounds from $l_{i} \rightarrow l_{k}\gamma$}
\begin{figure}[!t]
\centering
\includegraphics[width = 0.4\textwidth]{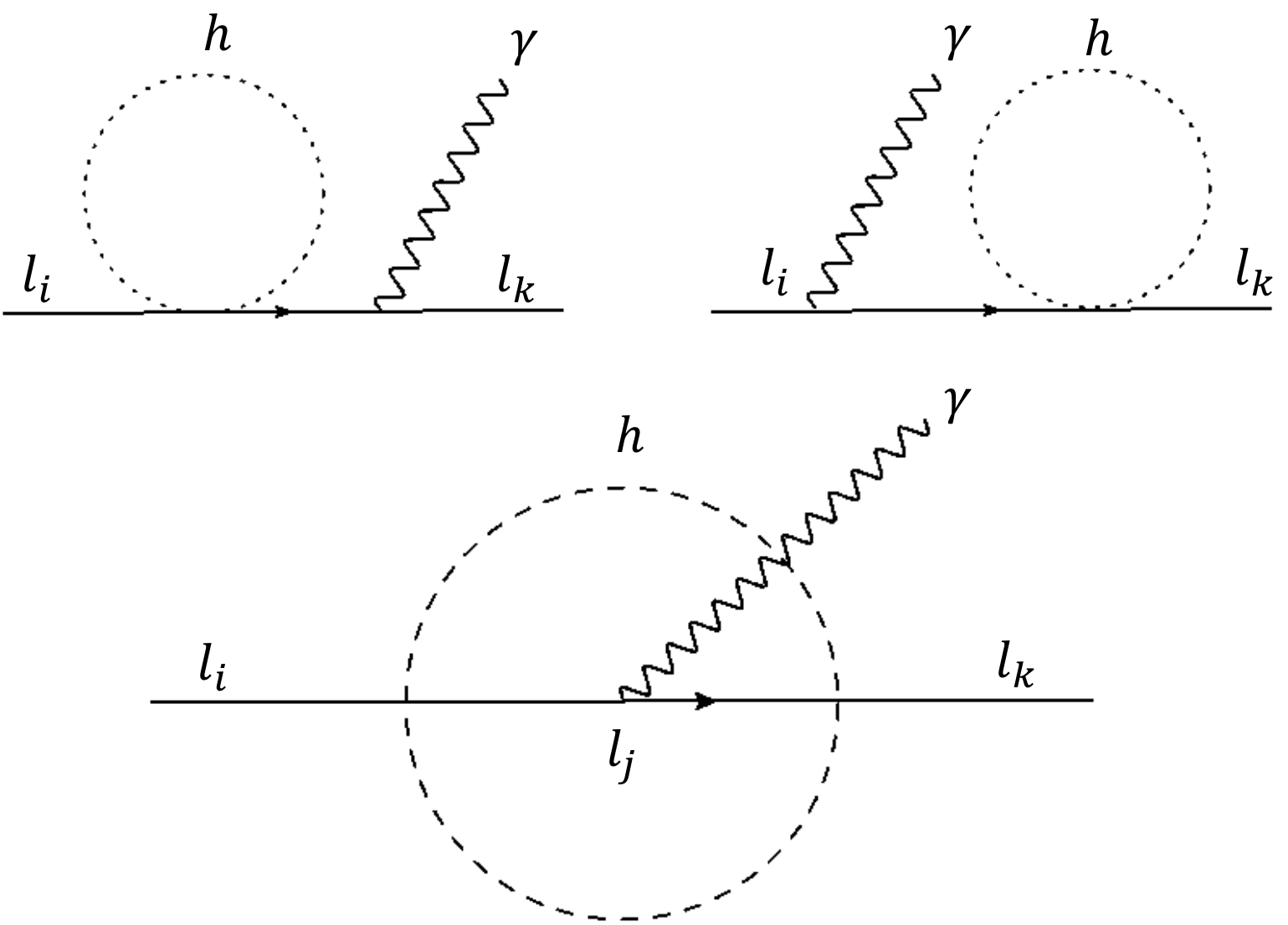}
\caption{\small FV decays $l_{i} \rightarrow \l_{k} \gamma$ through one (top) and two loops (bottom). }
\label{fig2}
\end{figure}

Stringent constraints can be obtained from the bounds on the FV decays $\tau \rightarrow \mu \gamma$, $\tau \rightarrow e\gamma$ and $\mu \rightarrow e \gamma$. The Feynman diagrams of these processes are shown in Figure \ref{fig2}. The 1-loop contributions are shown on the top row of the figure, where the photon could be emitted from the initial or final state lepton. The two contributions cancel one another and the contribution at one loop vanishes. Thus the leading contribution arises at 2-loops \footnote{Notice that there are two more 2-loop diagrams where the photon is emitted from the initial and final states, however, these two contribution cancel each other in exactly the same manner as in the 1-loop case.}

Calculating the 2-loop diagram is somewhat subtle and we show the details in Appendix \ref{appendix2}. For each decay process, the structure of the matrix element and the corresponding Wilson coefficients depend on the lepton inside the loop, i.e., each decay will have 3 contributions corresponding to setting the particle in the loop $j = \{\tau, \mu, e\}$. In order to set upper bounds on the Wilson coefficients, we isolate each contribution individually. This will lead to 9 different decay processes. For example, the decay width $\Gamma_{\tau \mu e}$ refers to the decay $\tau \rightarrow e\gamma$ with $\mu$ running in the loop.

Utilizing the results in Appendix \ref{appendix2}, assuming $m_{\tau} \gg m_{\mu} \gg m_{e}$, and setting the renormalization scale $\mu = m_{j}$, the decay widths are given by
\begin{align}
  \Gamma_{\mu\tau e} & \simeq \frac{\alpha |C_{\tau\mu}|^{2}|C_{\tau e}|^{2}}{16(4\pi)^{8}v^{4}} m_{\tau}^{2}m_{\mu}^{3} \Bigg[ \log{\Big( \frac{M_{h}^{2}}{m_{\tau}^{2}}\Big)}-\frac{\pi^{2}}{3}\Bigg]^{2},\label{eq:ltolgamma_1}\\
 \Gamma_{\mu\mu e} & \simeq \frac{\alpha |C_{\mu\mu}|^{2}|C_{\mu e}|^{2}}{9(4\pi)^{8}v^{4}} m_{\mu}^{5} \Bigg[ \log{\Big( \frac{M_{h}^{2}}{m_{\mu}^{2}}\Big)}-\frac{\pi^{2}}{4}\Bigg]^{2},\label{eq:ltolgamma_2}\\
  \Gamma_{\mu e e} & \simeq \frac{\alpha |C_{\mu e}|^{2}|C_{ee}|^{2}}{144(4\pi)^{8}v^{4}} m_{\mu}^{5} \log^{2}{\Big( \frac{M_{h}^{2}}{m_{e}^{2}}\Big)},\label{eq:ltolgamma_3}\\
   \Gamma_{\tau\tau\mu} & \simeq \frac{\alpha |C_{\tau\tau}|^{2}|C_{\tau\mu}|^{2}}{9(4\pi)^{8}v^{4}} m_{\tau}^{5} \Bigg[ \log{\Big( \frac{M_{h}^{2}}{m_{\tau}^{2}}\Big)}-\frac{\pi^{2}}{4}\Bigg]^{2},\label{eq:ltolgamma_4}\\
  \Gamma_{\tau \mu\mu} & \simeq \frac{\alpha |C_{\tau \mu}|^{2}|C_{\mu\mu}|^{2}}{144(4\pi)^{8}v^{4}} m_{\tau}^{5} \log^{2}{\Big( \frac{M_{h}^{2}}{m_{\mu}^{2}}\Big)},\label{eq:ltolgamma_5}\\
  \Gamma_{\tau e \mu} & \simeq \frac{\alpha |C_{\tau e}|^{2}|C_{\mu e}|^{2}}{144(4\pi)^{8}v^{4}} m_{\tau}^{5} \log^{2}{\Big( \frac{M_{h}^{2}}{m_{e}^{2}}\Big)},\label{eq:ltolgamma_6}\\
   \Gamma_{\tau\tau e} & \simeq \frac{\alpha |C_{\tau\tau}|^{2}|C_{\tau e}|^{2}}{9(4\pi)^{8}v^{4}} m_{\tau}^{5} \Bigg[ \log{\Big( \frac{M_{h}^{2}}{m_{\tau}^{2}}\Big)}-\frac{\pi^{2}}{4}\Bigg]^{2},\label{eq:ltolgamma_7}\\
   \Gamma_{\tau \mu e} & \simeq \frac{\alpha |C_{\tau \mu}|^{2}|C_{\mu e}|^{2}}{144(4\pi)^{8}v^{4}} m_{\tau}^{5} \log^{2}{\Big( \frac{M_{h}^{2}}{m_{\mu}^{2}}\Big)},\label{eq:ltolgamma_8}\\
  \Gamma_{\tau e e} & \simeq \frac{\alpha |C_{\tau e}|^{2}|C_{e e}|^{2}}{144(4\pi)^{8}v^{4}} m_{\tau}^{5} \log^{2}{\Big( \frac{M_{h}^{2}}{m_{e}^{2}}\Big)}\label{eq:ltolgamma_9}.
\end{align}

The experimental limits $\text{Br}(\mu \rightarrow e\gamma)  < 4.2 \times 10^{-13}$ \cite{ParticleDataGroup:2012pjm} can be used in the decays (\ref{eq:ltolgamma_1}), (\ref{eq:ltolgamma_2}) and (\ref{eq:ltolgamma_3}). The decay $\Gamma_{\mu\tau e}$ yields bounds $|C_{\tau \mu}| |C_{\tau e}| < 7.83 \times 10^{-5}$, whereas $\Gamma_{\mu\mu e}$ yields $|C_{\mu \mu}| |C_{\mu e}| < 4.4 \times 10^{-4}$, and $\Gamma_{\mu e e}$ translates to  $|C_{\mu e}| |C_{e e}| < 8.28 \times 10^{-4}$. On the other hand, the limit $\text{Br}(\tau \rightarrow \mu \gamma)  < 4.4 \times 10^{-8}$ \cite{ParticleDataGroup:2012pjm} can be used in the decays (\ref{eq:ltolgamma_4}), (\ref{eq:ltolgamma_5}) and (\ref{eq:ltolgamma_6}), with the decay $\Gamma_{\tau\tau\mu}$ leading to the bound $|C_{\tau\tau}||C_{\tau \mu}|< 0.66$ and the decay $\Gamma_{\tau\mu\mu}$ leading to the bound $|C_{\tau\mu}||C_{\mu \mu}|< 1.12$, whereas the decay $\Gamma_{\tau e\mu}$ leads to the bound $|C_{\tau e}||C_{\mu e}|< 0.64$. Finally, the experimental limits $\text{Br}(\tau \rightarrow e\gamma) < 3.3 \times 10^{-8}$ is used in last 3 decays in Eqs. (\ref{eq:ltolgamma_7}), (\ref{eq:ltolgamma_8}) and (\ref{eq:ltolgamma_9}), with $\Gamma_{\tau\tau e}$ yielding the bound $|C_{\tau\tau}||C_{\tau e}|<0.57$,  $\Gamma_{\tau\mu e}$ yielding the bound $|C_{\tau\mu}||C_{\mu e}|<0.97$ and finally  $\Gamma_{\tau e e}$ yielding the bound $|C_{\tau e}||C_{e e}|<0.55$.

Notice that bounds obtained here are roughly an order of magnitude weaker than the bounds obtained from $l \rightarrow 3l'$ decays. The reason for this is that the former case proceeds through two loops, whereas the latter proceeds through one loop.

As the case with the decays $l \rightarrow l_{1}l_{2}l_{3}$, the Belle II experiment is projected to provide stronger bounds \cite{Aushev:2010bq, Belle-II:2022cgf, Calibbi:2017uvl,Calibbi:2017uvl}, with projected branching ratios that are about an order of magnitude stronger than the current limits. For example, the projected Belle II constraints on the decay $\tau \rightarrow \mu\gamma$ are $\text{Br}(\tau \rightarrow \mu\gamma) < 1 \times 10^{-9}$. This can be used in $\Gamma_{\tau\tau\mu}$, $\Gamma_{\tau\mu\mu}$ and $\Gamma_{\tau e\mu}$ to yield the projections $|C_{\tau\tau}||C_{\tau \mu}| < 9.92 \times 10^{-2}$, $|C_{\tau\mu}||C_{\mu \mu}| < 0.17$ and $|C_{\tau e}||C_{\mu e}| < 9.66 \times 10^{-2}$, respectively. The projected limits are summarized in Table \ref{table1}. 

\subsection{Constraints from muonium-antimuonium oscillations}\label{sec:Muon_Osc}
\begin{figure}[!t]
\centering
\includegraphics[width = 0.5\textwidth]{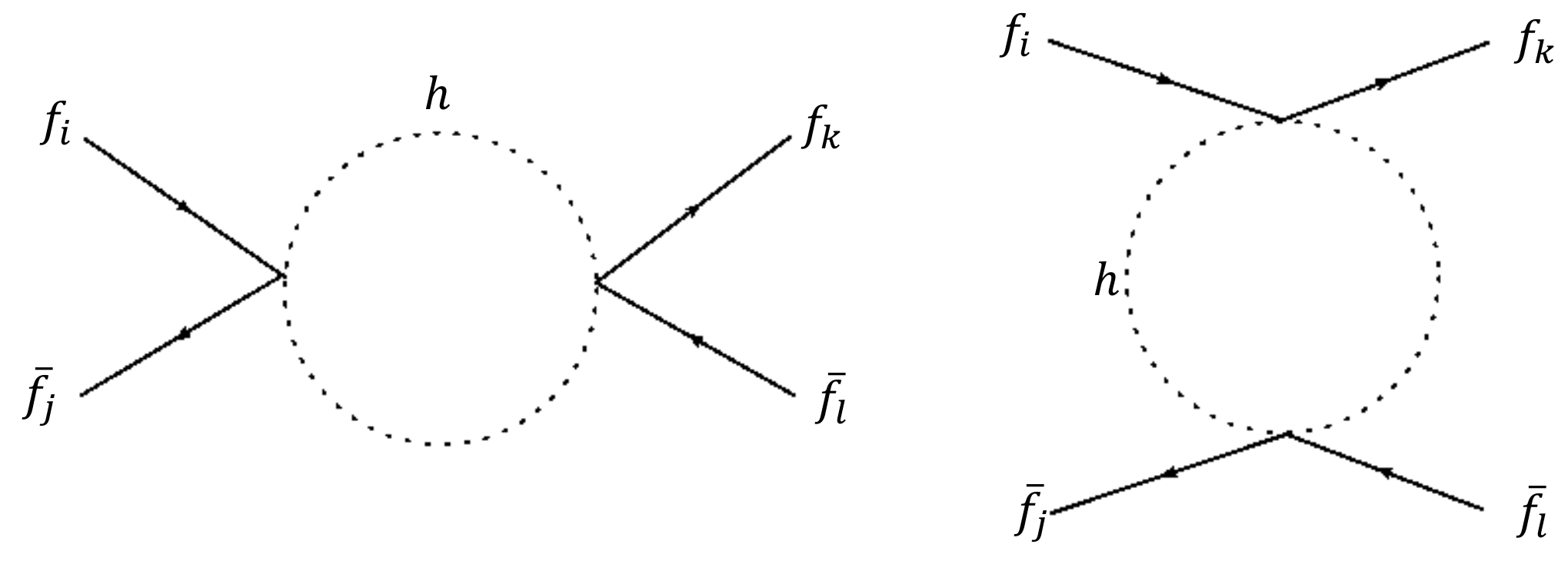}
\caption{\small The $f_{i}\overline{f}_{j} \rightarrow f_{k}\overline{f}_{l}$ scattering through the $h^{2}ff$ couplings. The left diagram is the s-channel, whereas the right diagram is the t-channel.}
\label{fig3}
\end{figure}

$\mu^{+}$ and $e^{-}$ can form a bound state called muonium. This bound state can oscillate to antimuonium $\mu^{-}e^{+}$ through the diagrams shown in Figure \ref{fig3}, with $f_{i} = e^{-}$, $\overline{f}_{j} = \mu^{+}$, $f_{k} = \mu^{-}$ and $\overline{f}_{l} = e^{+}$. The time-integrated $M-\overline{M}$ conversion probability is constrained by the MACS experiment at PSI \cite{Willmann:1998gd}
\begin{equation}\label{eq:muonium_bound}
    P(M \rightarrow \overline{M}) < 8.3 \times 10^{-11}/ S_{B},
\end{equation}
where $S_{B}$ accounts for the splitting of muonium in the magnetic field of the detectors and is given by $S_{B} = 0.35$ for $(S\pm P) \times (S\pm P)$ operators and $S_{B} = 0.9$ for $P \times P$ operators. In this paper, we chose to be conservative and set $S_{B} = 0.35$.
The loops in the s- and t-channels in Figure \ref{fig3} are given by Eq. \ref{eq:fftoff2}, which can be integrated out in the non-relativistic limit, yielding the following effective Lagrangian 
\begin{equation}\label{eq:muonium_eff}
    \mathcal{L}_{\text{eff}} = \frac{C_{\mu e}^{2}}{32\pi^{2}v^{2}} \log{\Big( \frac{m_{\mu}^{2}}{M_{h}^{2}}\Big)}[\overline{\mu}e][\overline{e}\mu],
\end{equation}
where we have set the renormalization scale $\mu^{2} = m_{\mu}^{2}$. The theoretical prediction for the conversion rate is governed by the matrix element
\begin{equation}\label{eq:muonium_conv1}
\mathcal{M}_{M} = \bra{\uparrow_{\mu}\downarrow_{\overline{e}}-\downarrow_{\mu}\uparrow_{\overline{e}}} \frac{1}{2}\mathcal{L}_{\text{eff}} \ket{\uparrow_{e}\downarrow_{\overline{\mu}}-\downarrow_{e}\uparrow_{\overline{\mu}}},
\end{equation}
where the factor of $1/2$ arises from the normalization of the initial and final states. Following the argument in \cite{Clark:2003tv}, the mass splitting between two states is given by
\begin{equation}\label{eq:mass_split1}
    |\Delta M| = 2|\mathcal{M}_{M}| = \frac{1}{\tau_{M}},
\end{equation}
where ${\tau_{M}}$ is the muonium oscillation time. A non-relativistic reduction of the effective Lagrangian in Eq. \ref{eq:muonium_eff} yields the following effective potential in position space
\begin{equation}\label{eq:muon_eff_V}
    V_{\text{eff}}(\vec{r}) =  \frac{C_{\mu e}^{2}}{64\pi^{2}v^{2}} \log{\Big( \frac{m_{\mu}^{2}}{M_{h}^{2}}\Big)} \delta^{3}(\vec{r}).
\end{equation}

We can assume that both $M$ and $\overline{M}$ are in the Coulombic ground state, such that their wavefunctions are $\phi_{100} = \exp(-\vec{r}/a_{M})/\sqrt{\pi a_{M}^{3}}$, with $a_{M} = 1/\alpha m_{\text{red}}$ being the muonium Bohr radius, and $m_{\text{red}} = m_{\mu}m_{e}/(m_{\mu}+m_{e}) \simeq m_{e}$ being the muonium reduced mass. Therefore, the mass splitting can easily be calculated as
\begin{equation}\label{eq:mass_split1}
 |\Delta M| \simeq 2\int d^{3}\vec{r}\phi_{100}^{*}(\vec{r}) V_{\text{eff}}(\vec{r}) \phi_{100}(\vec{r}) \simeq  \frac{C_{\mu e}^{2}}{32 \pi^{3}v^{2} a_{M}^{3}} \log{\Big( \frac{m_{\mu}^{2}}{M_{h}^{2}}\Big)},
\end{equation}
and the conversion rate readily follows
\begin{equation}\label{eq:muonium_conv1}
 P(M \rightarrow \overline{M}) = \int_{0}^{\infty}dt\Gamma_{\mu} \sin^{2}{(\Delta M t)}e^{-\Gamma_{\mu}t} = \frac{2}{4+\Gamma_{\mu}^{2}/(\Delta M)^{2}}.
\end{equation}

Given the bound in Eq. (\ref{eq:muonium_bound}), we find the upper limit on $|C_{\mu e}| < 0.39$.

\subsection{Constraints from the magnetic dipole moment and the $g-2$ anomaly} \label{sec:g-2}
It was first shown by the E821 experiment at BNL \cite{Muong-2:2006rrc} and later confirmed by the E989 experiment at Fermilab \cite{Muong-2:2021ojo, Muong-2:2021ovs, Muong-2:2021vma}, that there is a discrepancy between the measured and predicted \cite{Aoyama:2020ynm} magnetic dipole moment of the muon. This discrepancy, known as the $g-2$ anomaly, currently stands at
\begin{equation}\label{eq:g-2_value}
    \Delta a_{\mu} = a_{\mu}^{\text{Exp}} - a_{\mu}^{\text{SM}} = (251 \pm 59) \times 10^{-11},
\end{equation}
with a significance of $4.2\sigma$. On the other hand, several lattice QCD groups have recently reported higher theoretical predictions compared to the data-driven approach and seem to agree with experiment \cite{Borsanyi:2020mff, Ce:2022kxy, Alexandrou:2022amy}. For the purposes of extracting the relevant bounds, we shall assume that the $g-2$ anomaly exists and that it is given by Eq. (\ref{eq:g-2_value}) above, and if future studies show that indeed the theory and experiment agree, then the bounds are simply ignored.
\begin{figure}[!t]
\centering
\includegraphics[width = 0.55\textwidth]{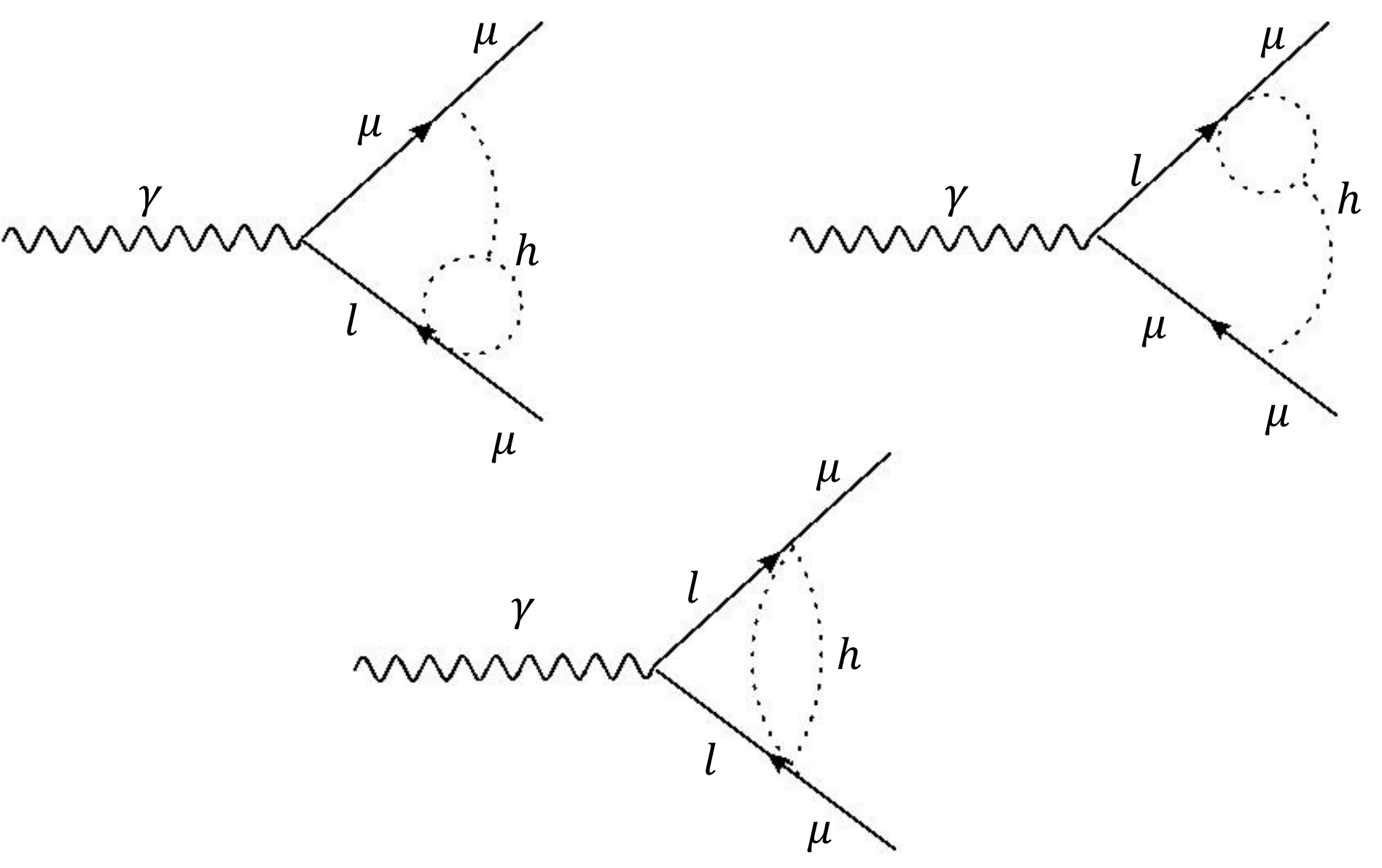}
\caption{\small FV contribution to the muon magnetic dipole moment through the couplings $h^{2}\overline{l}l$. Here, $l = \{e, \mu , \tau \}$.}
\label{fig4}
\end{figure}

The possibility of the effective coupling $h^{2}\overline{\mu}\mu$ solving the $g-2$ anomaly was considered in \cite{Abu-Ajamieh:2022nmt}, where it was shown that this type of coupling can accommodate the anomaly if this coupling is large enough. It was also shown that such a deviation from the SM would point to a scale of NP $\sim 10 -18$ TeV through unitarity arguments, which can be lowered to $\sim 5$ TeV if the Higgs couplings to $W/Z$ conform to the SM predictions.

Here we generalize the situation to FV $h^{2}\overline{l}l$ couplings. These couplings contribute to the muon magnetic dipole moment at 2 loops as shown in the diagrams in Figure \ref{fig4}. Notice that the FV case corresponds to $l = e, \tau$. These diagrams can be evaluated using the same techniques illustrated in the appendices and in \cite{Abu-Ajamieh:2022nmt}, and they are found to provide the following contribution to $(g-2)_{\mu}$
\begin{equation}\label{eq:g-2_contributions}
    \Delta a_{\mu}^{l} \simeq \frac{C_{\mu l}^{2}}{2(4\pi)^{4}v^{2}}m_{\mu}m_{l} \Bigg[2 \log^{2}\Big( \frac{m_{l}^{2}}{\Lambda^{2}}\Big)-\Big(1+\frac{2m_{\mu}}{3m_{l}}\Big)\log \Big( \frac{M_{h}^{2}}{\Lambda^{2}}\Big)+\frac{\pi^{2}}{3}\Bigg],
\end{equation}
 where the UV cutoff $\Lambda \gg M_{h}$. Setting $\Lambda = 10$ TeV, the $g-2$ anomaly in Eq. (\ref{eq:g-2_value}) can be explained with the following values\footnote{The coupling $C_{\mu\mu}$ defined here is rescaled compared to the coupling $C_{\mu 2}$ defined in \cite{Abu-Ajamieh:2022nmt}. The two couplings are related as follows: $C_{\mu 2} = \frac{v C_{\mu\mu}}{\sqrt{2}m_{\mu}}$. With this rescaling, both results are consistent.}
\begin{align}
|C_{\tau\mu}| & \simeq 0.26 \pm 0.03,  \label{eq:g-2_explanation1}\\
|C_{\mu\mu} | & \simeq 0.79 \pm 0.1,   \label{eq:g-2_explanation2}\\ 
|C_{\mu e} | & \simeq 6.34 \pm 0.8.  \label{eq:g-2_explanation3}
\end{align}
\subsection{Constraints form the electric dipole moment}\label{sec:EDM}
In general, FV coupling of the form $C_{ij}$ can contribute to the Electric Dipole Moment (EDM) of electrons and muon if such couplings are complex. In such case, the EDM will be proportional to the imaginary parts of $C_{ij}$, however, as we are assuming real couplings, there will be no constraints from the EDM of the electron or the muon.

\subsection{LEP constraints}\label{sec:lep}
Constraints can be obtained from LEP from the processes $e^{+}e^{-} \rightarrow \mu^{+}\mu^{-}$, $\tau^{+}\tau^{-}$. These processes are shown in Figure \ref{fig3}. The s-channel involves the couplings $C_{ee}$, $C_{\mu\mu}$ and $C_{\tau\tau}$ and thus does not lead to any FV. Therefore, we ignore it by setting these couplings to $0$. On the other hand, the t-channel involves the FV couplings $C_{\mu e}$ and $C_{\tau e}$. Details for calculating the loop are given in Appendix \ref{appendix3}. Using the explicit expression of the loop integral in Eq. (\ref{eq:fftoff3}), it is a simple exercise to calculate the cross-section of the above processes. Neglecting the masses of the initial and final state leptons, and using $\sqrt{s} = 207$ GeV\footnote{Although the COM energy of LEP is $209$ GeV, the relevant COM energy for the processes $e^{+}e^{-} \rightarrow \mu^{+}\mu^{-}$, $\tau^{+}\tau^{-}$ quoted in \cite{ALEPH:2006bhb} is actually $207$ GeV.} and a UV cutoff $\Lambda = 10^{4}$ GeV, we find
\begin{equation}\label{eq:LEP}
    \sigma(e^{+}e^{-} \rightarrow \mu^{+}\mu^{-}(\tau^{+}\tau^{-})) \simeq 2.2 \times 10^{-2} C_{\mu (\tau) e}^{4} \hspace{2 mm} \text{fb}.
\end{equation}

The $1\sigma$ uncertainties on $\sigma(e^{+}e^{-} \rightarrow \mu^{+}\mu^{-}(\tau^{+}\tau^{-}))$ are given by $0.088$ ($0.11$) pb \cite{ALEPH:2006bhb}, which can be translated into the rather weak bounds $|C_{\mu e}| < 9$ ($|C_{\tau e}| < 9.52$). This is expected as these processes are proportional to four powers of the couplings and thus cannot compete with decay processes, which are proportional to only two powers of the coupling. This is consistent with the case of FV from Yukawa couplings, see for instance \cite{Harnik:2012pb}.

\subsection{Constraints from $\mu$ conversion in nuclei}\label{sec:Conversion}
\begin{figure}[!t]
\centering
\includegraphics[width = 0.55\textwidth]{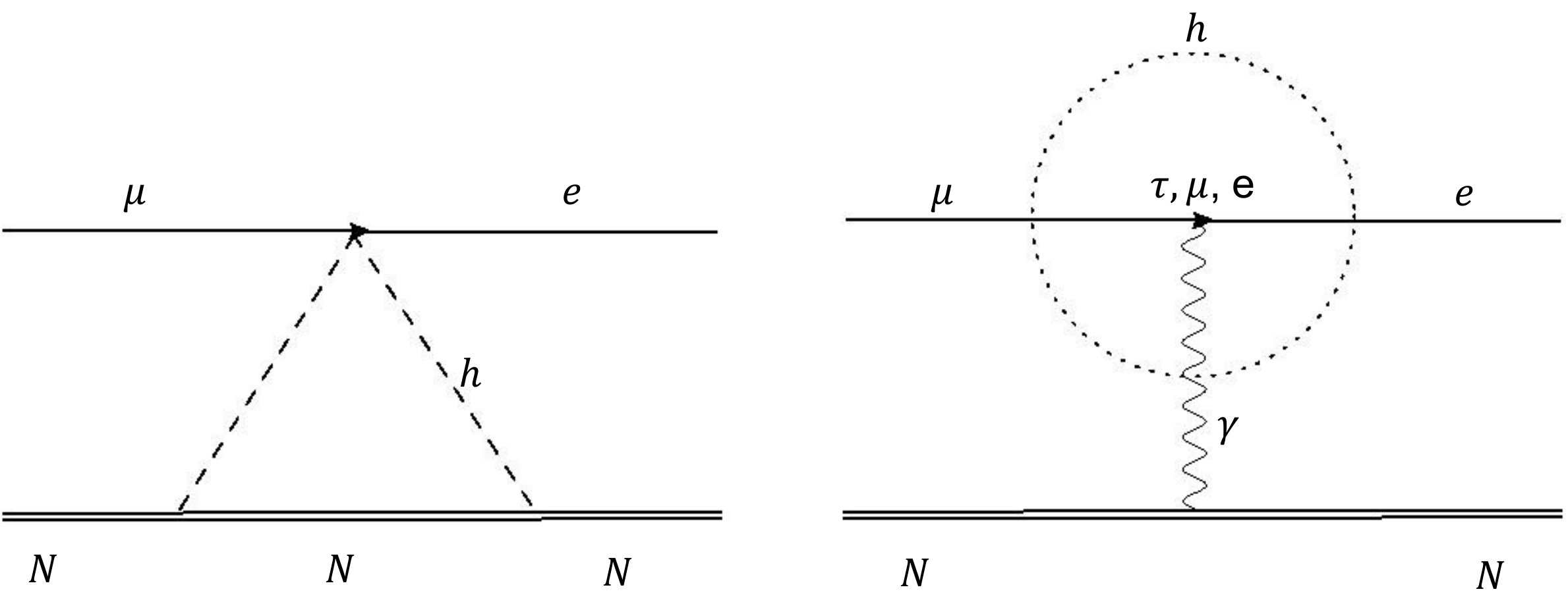}
\caption{\small FV contribution to $\mu \rightarrow e$ conversion in nuclei.}
\label{fig5}
\end{figure}

The experimental searches for the conversion of $\mu \rightarrow e$ in nuclei can be used to set limits of the leptonic effective FV couplings $h^{2}\overline{l}l$. This process can proceed at one and two loops as shown in Figure \ref{fig5}. In the notation of \cite{Kitano:2002mt}, the diagram on the left (right) is called the scalar (tensor) contribution. 
The scalar contribution can set limits on the coupling $C_{\mu e}$. On the other hand, the tensor diagram can provide constraints on the combinations $|C_{\tau\mu}||C_{\tau e}|$, $|C_{\mu\mu}||C_{\mu e}|$ or $|C_{\mu e}||C_{ee}|$ depending on the lepton running inside the loop. However, the tensor contribution is not expected to compete with the bounds from $l_{i} \rightarrow l_{k}\gamma$ and therefore we neglect it here. We present the detailed calculation in Appendix \ref{appendix4}. 

From Eq. (\ref{eq:mu2e_bound}), the bound $\text{Br}(\mu \rightarrow e) < 7 \times 10^{-13}$ $@$ $90\%$ C.L. \cite{SINDRUMII:2006dvw} translates into the upper bound $|C_{\mu e}| < 0.34$. On the other hand, the Mu2e experiment is planning on improving this limit to $\text{Br}(\mu \rightarrow e)< 10^{-16}$ \cite{Kargiantoulakis:2019rjm}. This would better the bound to become $|C_{\mu e}| < 4.56 \times 10^{-3}$.

\subsection{Higgs FV decays}
\begin{figure}[!t]
\centering
\includegraphics[width = 0.40\textwidth]{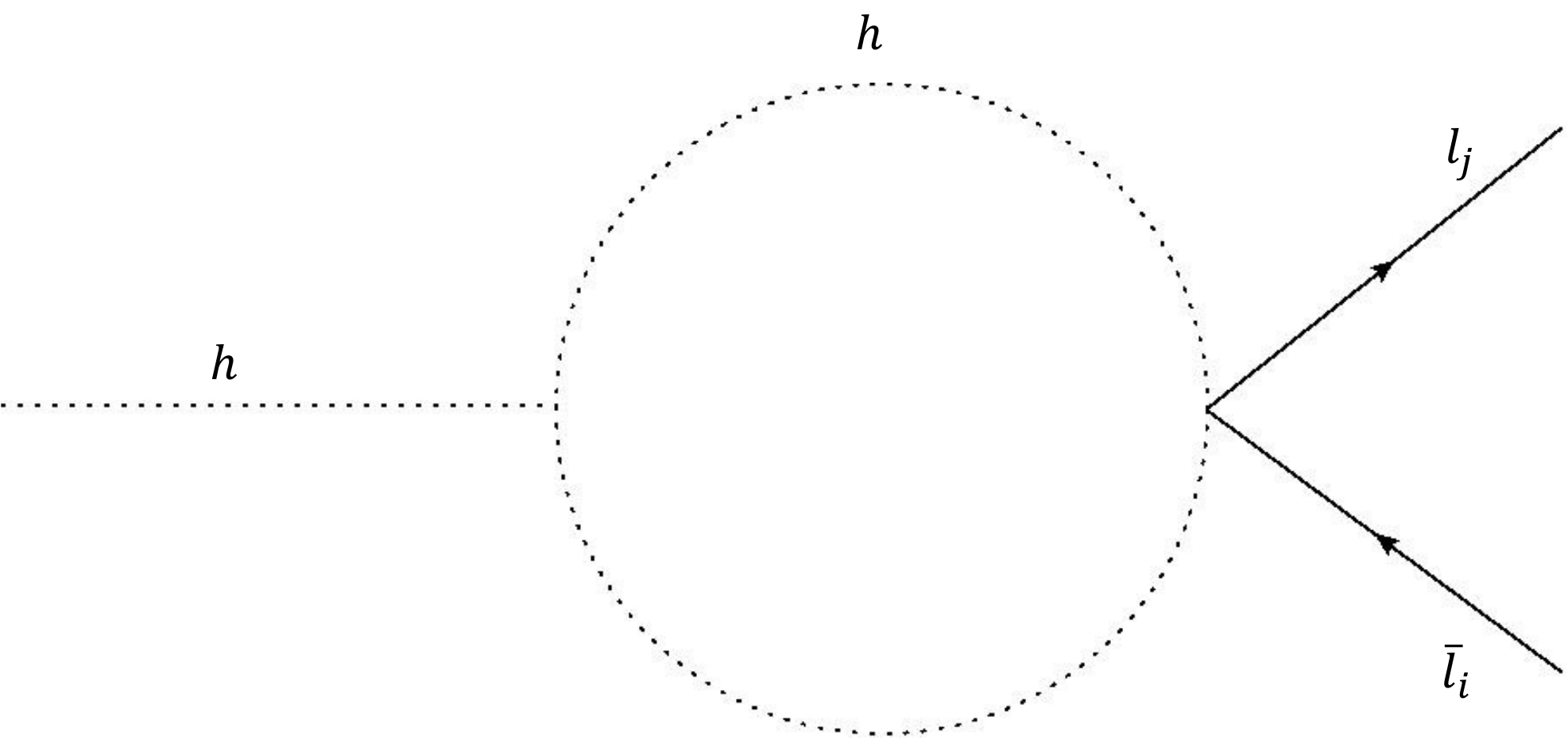}
\caption{\small FV Higgs decays to $l_{i}l_{j}$.}
\label{fig6}
\end{figure}

Higgs FV decays can be used to set constraints on the leptonic couplings $C_{ij}$. These decays proceed at one loop as shown in Figure \ref{fig6}. The diagram is easily evaluated using Dimensional Regularization (DR), and the decay width is given by
\begin{equation}\label{eq:H_decay}
    \Gamma(h \rightarrow \overline{l}_{i}l_{j}) \simeq \frac{9C_{ij}^{2}M_{h}^{5}}{4(4\pi)^{5}v^{4}},
\end{equation}
where we have set the renormalization scale $\mu^{2} = M_{h}^{2}e^{-\pi/\sqrt{3}}$ and neglected the masses of the final state. The latest bounds on these decays can be obtained from \cite{ParticleDataGroup:2018ovx}. In specific, we have the following bounds:\footnote{The quoted bounds are $@$ $95\%$ CL. Therefore, we rescale them to $90\%$ to be consistent with the other bounds. } $\text{Br}(h \rightarrow \mu e) < 3.5 \times 10^{-4}$, $\text{Br}(h \rightarrow \tau e) < 6.1 \times 10^{-3}$ and $\text{Br}(h \rightarrow \tau\mu) < 2.5 \times 10^{-3}$. These bounds translate into the constraints: $|C_{\mu e}| < 0.25$, $|C_{\tau e}| < 1.04$ and $|C_{\tau \mu}| < 0.67$ respectively. For completeness, \cite{ParticleDataGroup:2018ovx} also provides the upper bound $\text{Br}(h \rightarrow e^{+}e^{-}) < 1.9 \times 10^{-3}$, which provides the constraint $|C_{ee}| < 0.58$.

The High-Luminosity (HL) LHC is expected to yield stronger bounds on the Higgs FV decays \cite{Banerjee:2016foh}. The projected limited on the decay $h \rightarrow e\mu$ is $3 \times 10^{-5}$, which translates into a projected bound of $|C_{\mu e}|< 7.3 \times 10^{-2}$. On the other hand, the project limit on the decays $h \rightarrow \mu\tau, \tau e$ is $3 \times 10^{-4}$, which leads to the upper bound of $|C_{\tau \mu}|,|C_{\tau e}| <0.23$.

\subsection{Constraints from $Z \rightarrow \overline{l}l$}\label{Sec;Z_decay1}
The excellent measurements of the $Z$ branching ratios suggest that they can be used to extract bounds on the leptonic FV couplings. The FV couplings $C_{ij}$ can contribute to the $Z$ decay through a process similar to the bottom diagram of Figure \ref{fig4}, with the photon being replaced with $Z$, and the external particles being leptons of the same flavor, whereas the internal leptons being of a different flavor. Using DR, we express the corrections to decay width of the $Z$ boson as
\begin{equation}\label{eq:Z_decay1}
    \delta \text{Br}(Z \rightarrow \overline{l}_{i}l_{i}) \simeq  \frac{(g_{V_{l}}^{2}+g_{A_{l}}^{2})M_{Z}}{48\Gamma_{Z}} \Big(  \frac{C_{ij}M_{h}}{16\pi^{2}v}  \Big)^{4} \Bigg[ \log^{2}\Big( \frac{M_{h}^{2}}{M_{Z}^{2}}\Big) -3 \log \Big( \frac{M_{h}^{2}}{M_{Z}^{2}}\Big) + \frac{\pi^{2}}{12} + \frac{7}{2}
    \Bigg]^{2},
\end{equation}
where $g_{V_{l}} = \frac{g}{2\cos{\theta_{W}}}(T_{l_{L}}^{3} -2 \sin^{2}{\theta_{W}}Q_{l})$ and $g_{A_{l}} = \frac{g}{2\cos{\theta_{W}}}(T_{l_{L}}^{3})$ are the vector and axial couplings of the lepton $l$ to the $Z$ boson in the standard notation, and $\Gamma_{Z} = 2.4952$ GeV is the measured decay width of the $Z$. The limits on non-FV leptonic $Z$ decays are given by \cite{ParticleDataGroup:2018ovx} 
\begin{align}
    \text{Br}(Z \rightarrow e^{+}e^{-}) & = (3363.2 \pm 4.2) \times 10^{-3} \%, \label{eq:Z_decay_bounds1_1}\\
     \text{Br}(Z \rightarrow \mu^{+}\mu^{-}) & = (3366.2 \pm 6.6) \times 10^{-3} \%, \label{eq:Z_decay_bounds1_2} \\
      \text{Br}(Z \rightarrow \tau^{+}\tau^{-}) & = (3369.8 \pm 8.3) \times 10^{-3} \% \label{eq:Z_decay_bounds1_3}.
\end{align}

Given these bounds, we can extract $90\%$ C.L. constraints on the FV couplings $C_{ij}$ by demanding that $ (\delta \text{Br})^{\text{FV}} < 1.645 (\delta \text{Br})^{\text{Exp}}$. Each bound can help constrain 3 different couplings depending on the flavor of the internal lepton, two of which are FV whereas one is flavor-conserving. Apart from the coupling $C_{ij}$, the correction in Eq. (\ref{eq:Z_decay1}) is identical for all lepton flavors. This means that for each decay mode, the upper limit for all three FV couplings will be identical.

The experimental limits in Eq. (\ref{eq:Z_decay_bounds1_1}) lead to the constraints $|C_{\tau e}|, |C_{\mu e}|, |C_{e e}| < 5.62$, whereas the limits in Eq. (\ref{eq:Z_decay_bounds1_2}) translate into constraints $|C_{\tau \mu}|, |C_{\mu \mu}|, |C_{\mu e}| < 7.04$ and the limits in Eq. (\ref{eq:Z_decay_bounds1_3}) yield the bounds $|C_{\tau \tau}|, |C_{\tau \mu}|, |C_{\tau e}| < 7.9$. These limits are comparable to the ones obtained from the LEP measurements above (see subsection \ref{sec:lep}), which is expected, as the experimental limits shown in Eqs. (\ref{eq:Z_decay_bounds1_1}) - (\ref{eq:Z_decay_bounds1_3}) are essentially obtained from LEP data. However, improved $Z$ decay measurements in future experiment, such as in the ILC \cite{Behnke:2013xla}; can improve the these limits through its proposed ultra-precision electroweak measurements.

\subsection{Constraints from $Z \rightarrow \overline{l}_{i}l_{j}$}\label{Sec;Z_decay2}
\begin{figure}[!t]
\centering
\includegraphics[width = 0.55\textwidth]{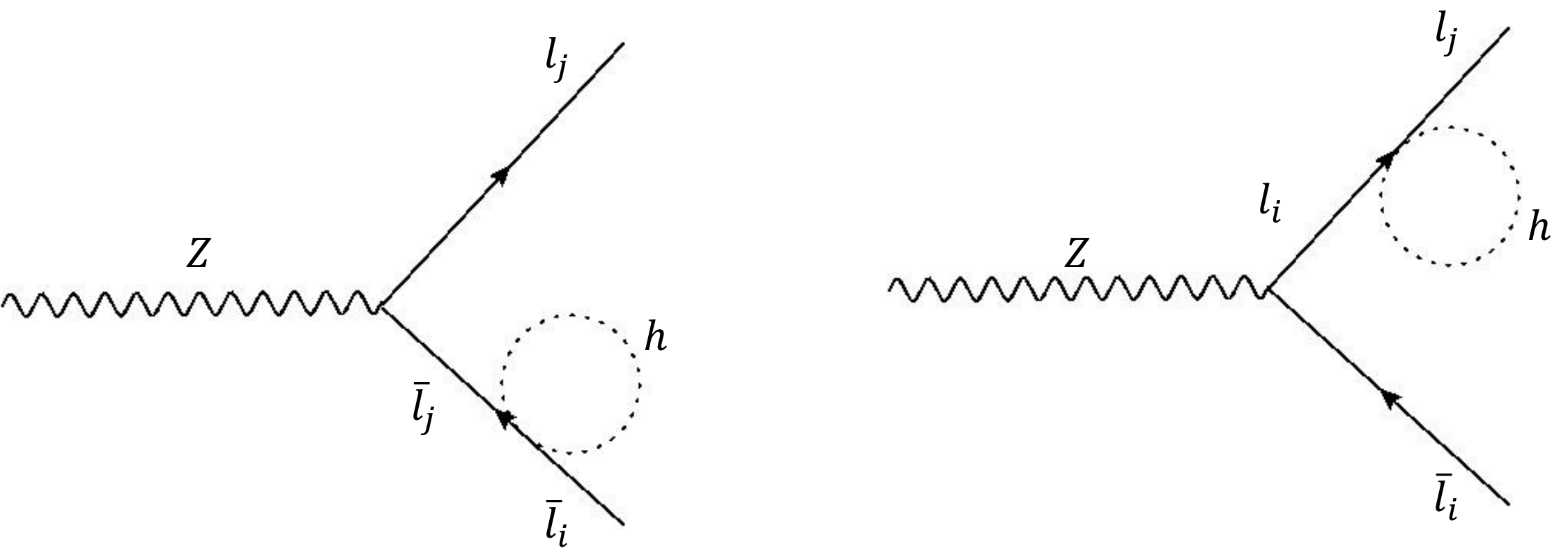}
\caption{\small FV $Z$ decays to $\overline{l}_{i}l_{j}$.}
\label{fig7}
\end{figure}

Better constraints can be obtained from the bounds on the FV decays of the $Z$ boson, because unlike the decays $Z \rightarrow \overline{l}l$ which starts at 2 loops, the decays $Z \rightarrow \overline{l}_{i}l_{j}$ start at 1 loop as shown in Figure \ref{fig7}. In addition, the experimental bounds on FV final states are more stringent compared to the flavor-conserving ones.

The corrections of the diagrams in Figure \ref{fig7} are easy to calculate by first integrating out the Higgs loop then calculating the tree-level diagram. Using DR in the $\overline{\text{MS}}$ scheme, and setting the renormalization scale $\mu = M_{h}$, we obtain
\begin{equation}\label{eq:Z_decay2}
    \delta \Gamma(Z \rightarrow \overline{l}_{i}l_{j}) \simeq \frac{C_{ij}^{2}(g_{V_{l}}^{2}+g_{A_{l}}^{2})}{6(4\pi)^{5}v^{2}} \frac{m_{i}^{2}M_{Z}M_{h}^{4}}{(m_{i}^{2}-m_{j}^{2})^{2}}.
\end{equation}

The limits on FV leptonic $Z$ decays are given by \cite{ParticleDataGroup:2018ovx}\footnote{Here too, the bounds quoted are @ $95\%$ C.L., and we rescale them to be @ $90\%$ C.L. to be consistent with the previous results.}
\begin{align}
    \text{Br}(Z \rightarrow e^{\pm}\mu^{\mp}) & = 7.5 \times 10^{-7}, \label{eq:Z_decay_bounds2_1}\\
     \text{Br}(Z \rightarrow e^{\pm}\tau^{\mp}) & = 9.8 \times 10^{-6}, \label{eq:Z_decay_bounds2_2} \\
      \text{Br}(Z \rightarrow \mu^{\pm}\tau^{\mp}) & = 1.2 \times 10^{-5}. \label{eq:Z_decay_bounds2_3}
\end{align}

Notice that for each decay, the corresponding $C_{ij}$ will have 2 possible upper limits depending on which particle is identified as $i$ and which one is identified as $j$. For example, for the first decay, we will have a different bound when we identify $i$ as $e$ and $j$ as $\mu$ compared to when these particles are flipped. As can clearly be seen from Eq. (\ref{eq:Z_decay2}), the strongest bound is obtained when $i$ is identified with the heavier of the two leptons. In the following, we quote the stronger of the two bounds. Specifically, Eqs. (\ref{eq:Z_decay_bounds2_1}), (\ref{eq:Z_decay_bounds2_2}) and (\ref{eq:Z_decay_bounds2_3}) lead to the constraints $|C_{\mu e}| < 1.59 \times 10^{-3}$, $|C_{\tau e}| < 9.65 \times 10^{-2}$ and $|C_{\tau \mu}| < 0.11$ respectively. 

\begin{figure}[!t]
\centering
\subfloat[$|C_{\mu\mu}|$ Vs. $|C_{\tau\mu}|$]{
  \includegraphics[width=54mm]{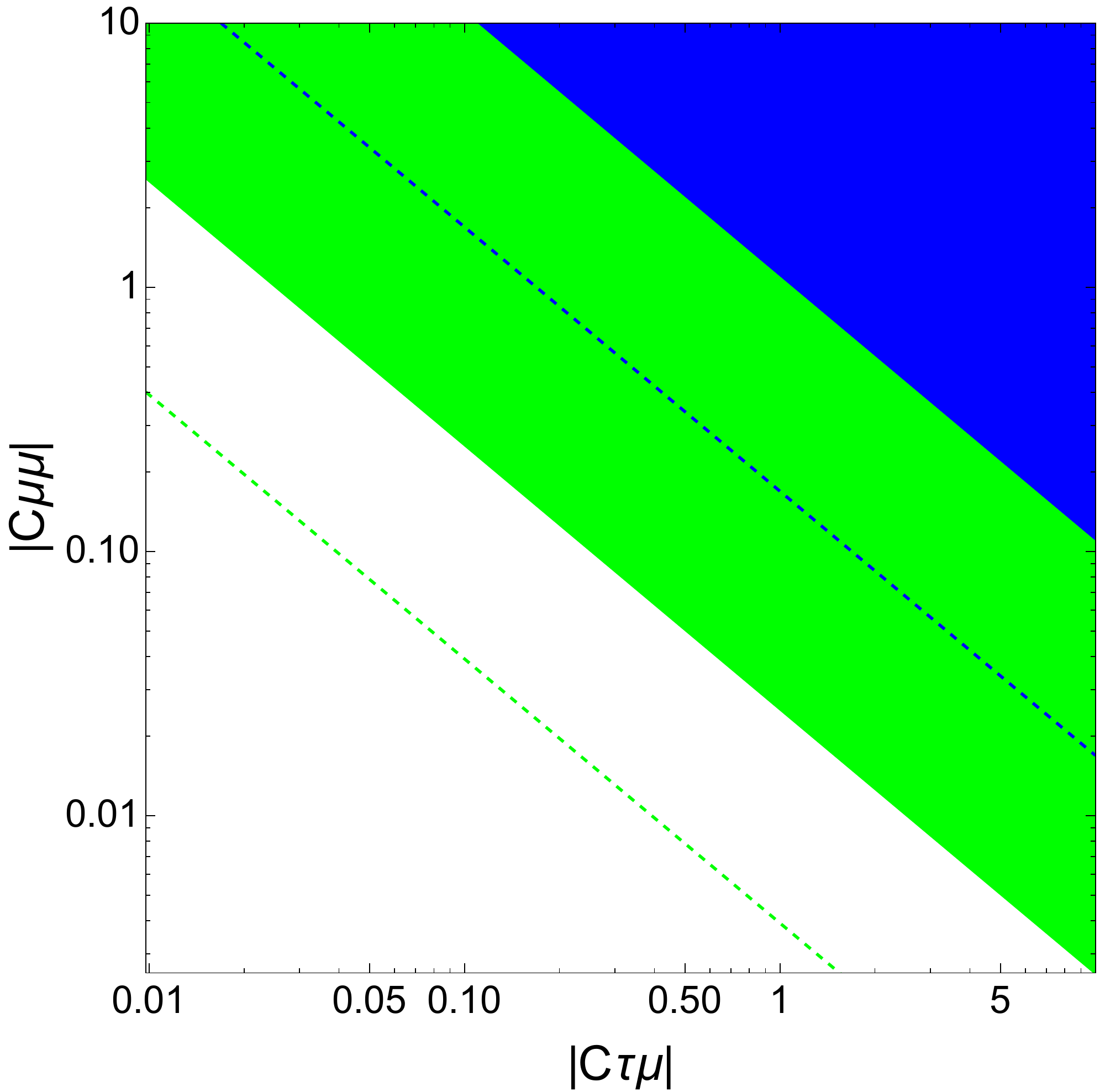}
}
\subfloat[$|C_{ee}|$ Vs. $|C_{\mu e}|$]{
  \includegraphics[width=54mm]{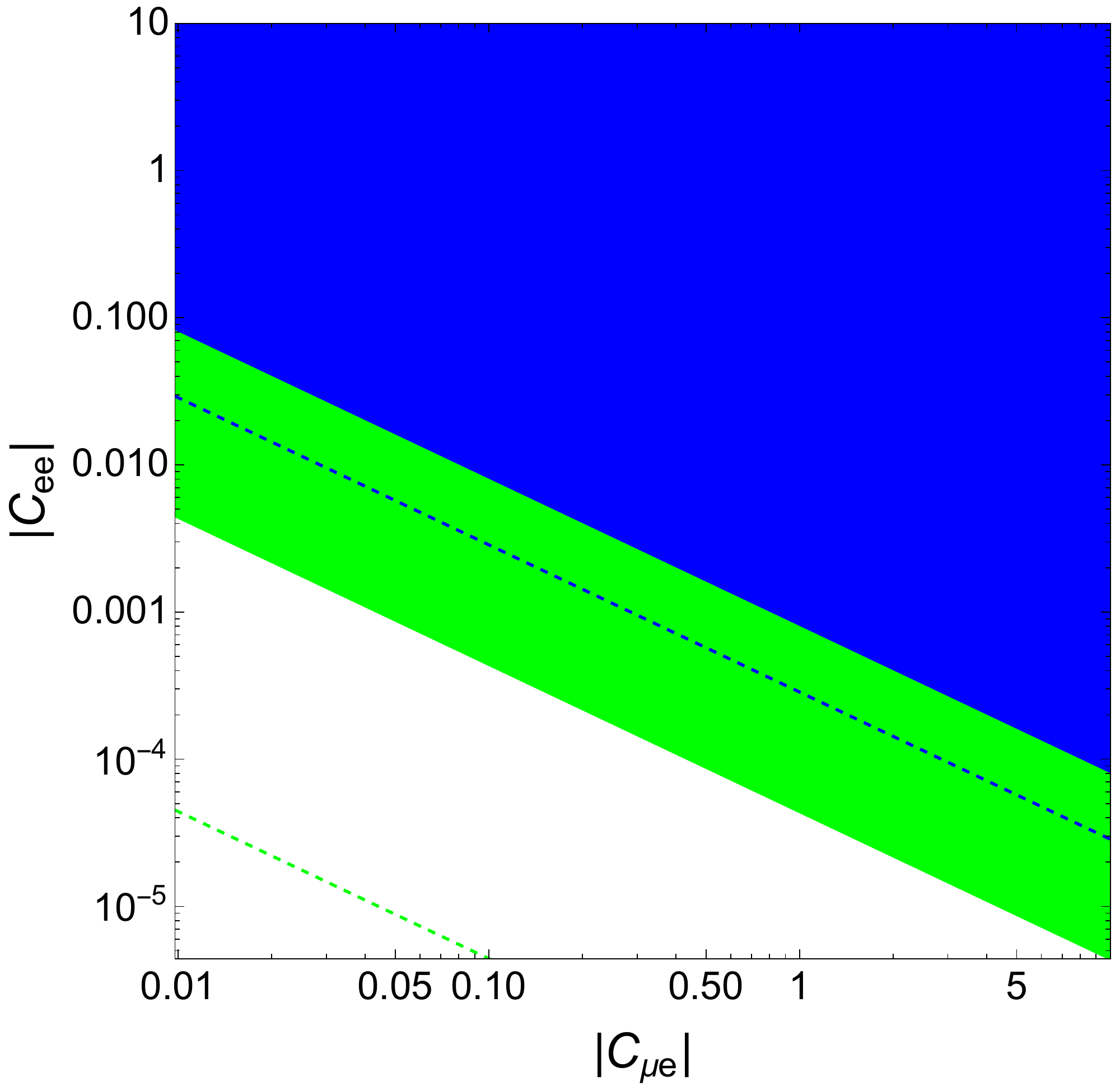}
}
\subfloat[$|C_{ee}|$ Vs. $|C_{\tau e}|$]{
  \includegraphics[width=54mm]{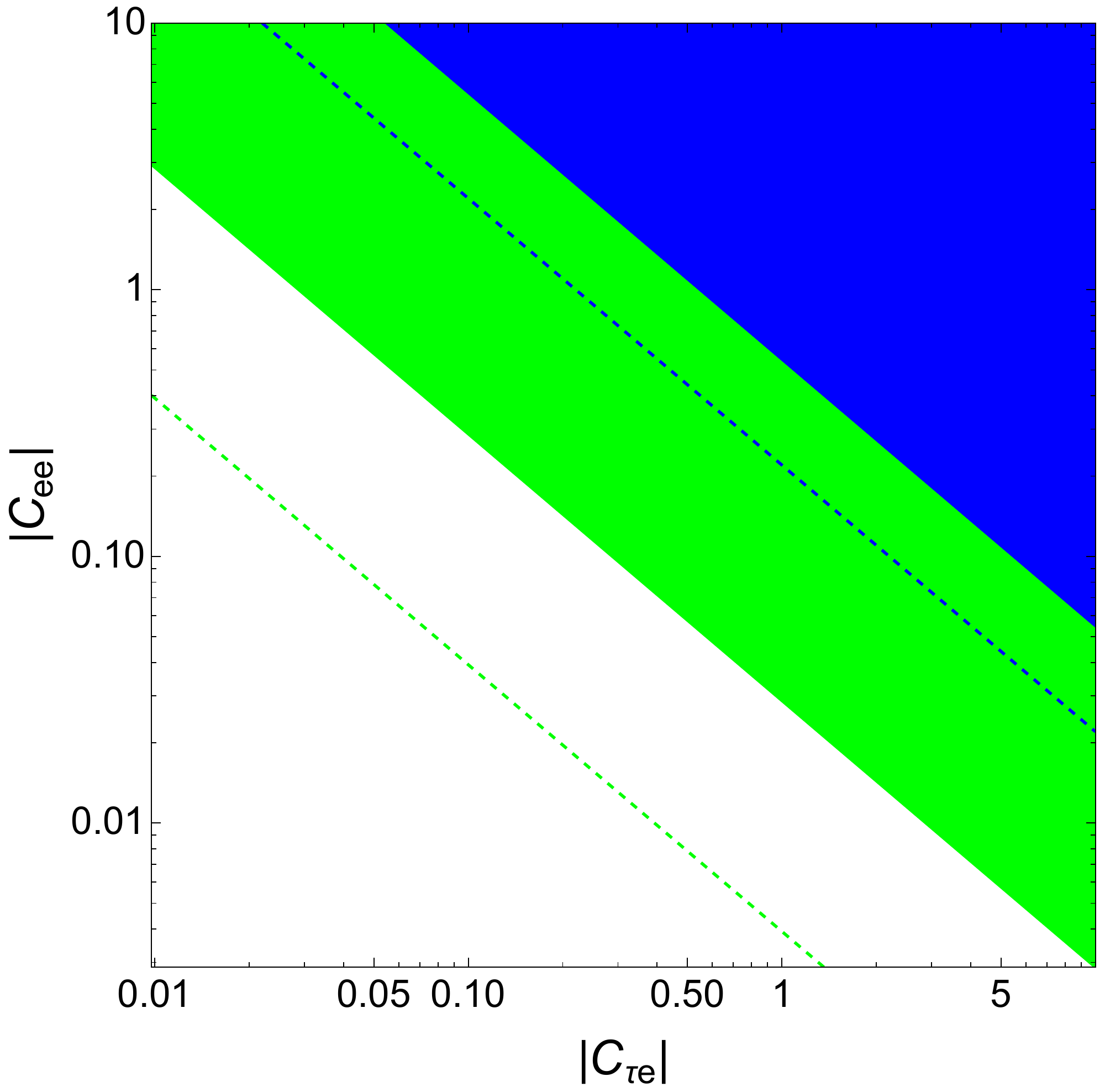}
}
\hspace{0mm}
\subfloat[$|C_{\mu e}|$ Vs. $|C_{\tau\mu}|$]{
  \includegraphics[width=54mm]{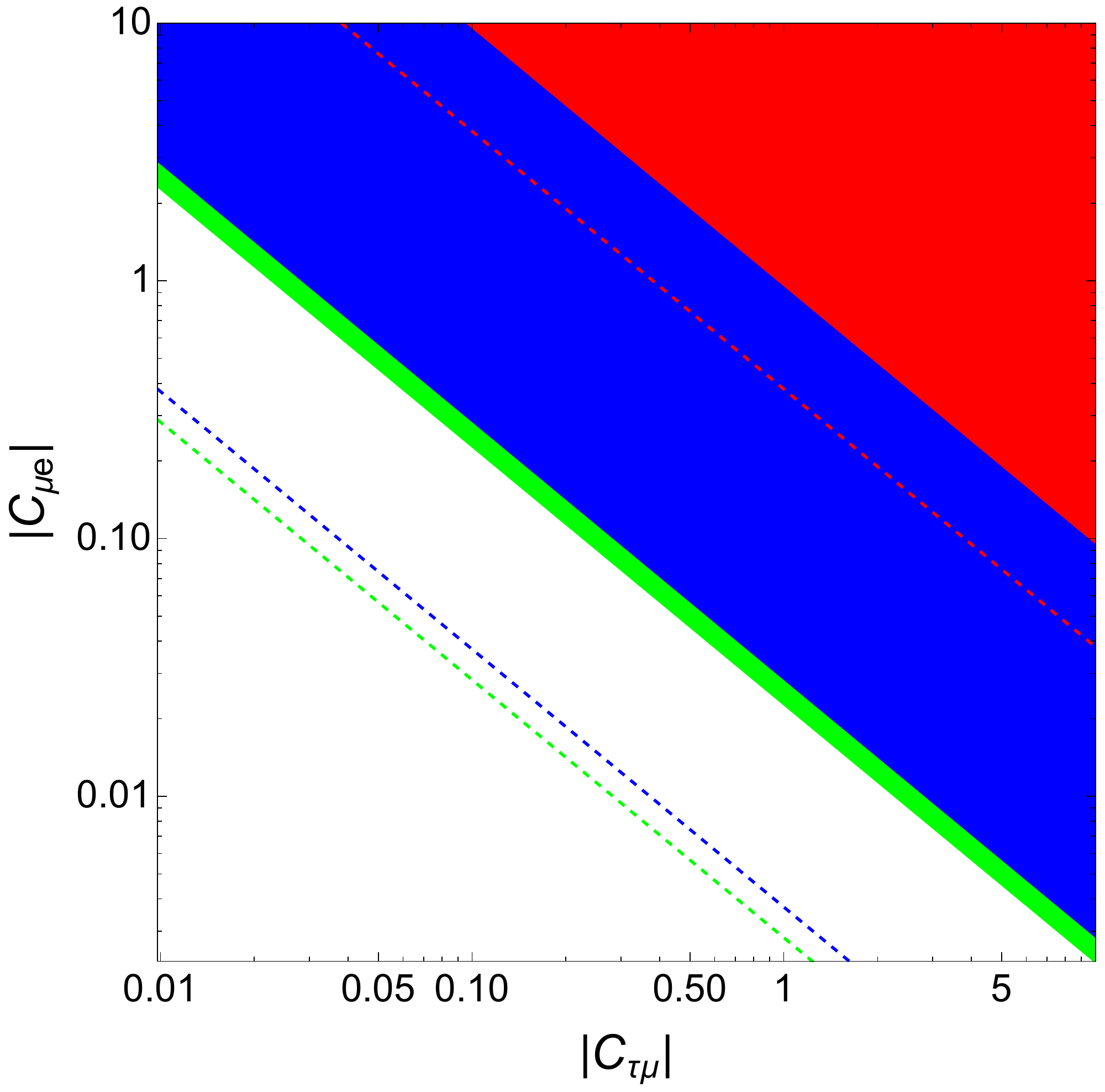}
}
\subfloat[$|C_{\mu e}|$ Vs. $|C_{\tau e}|$]{
  \includegraphics[width=54mm]{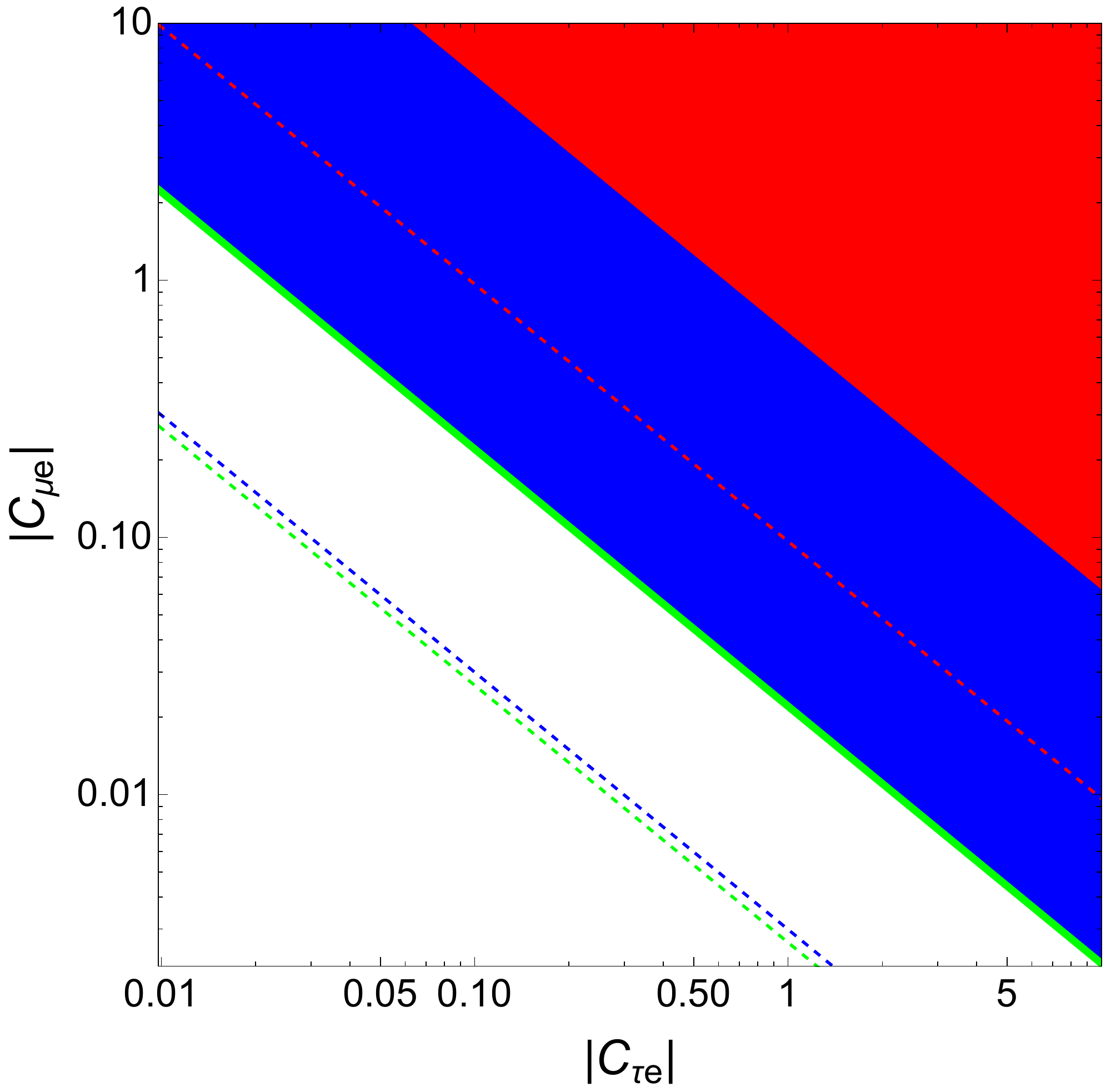}
}
\subfloat[$|C_{\mu\mu}|$ Vs. $|C_{\tau e}|$]{
  \includegraphics[width=54mm]{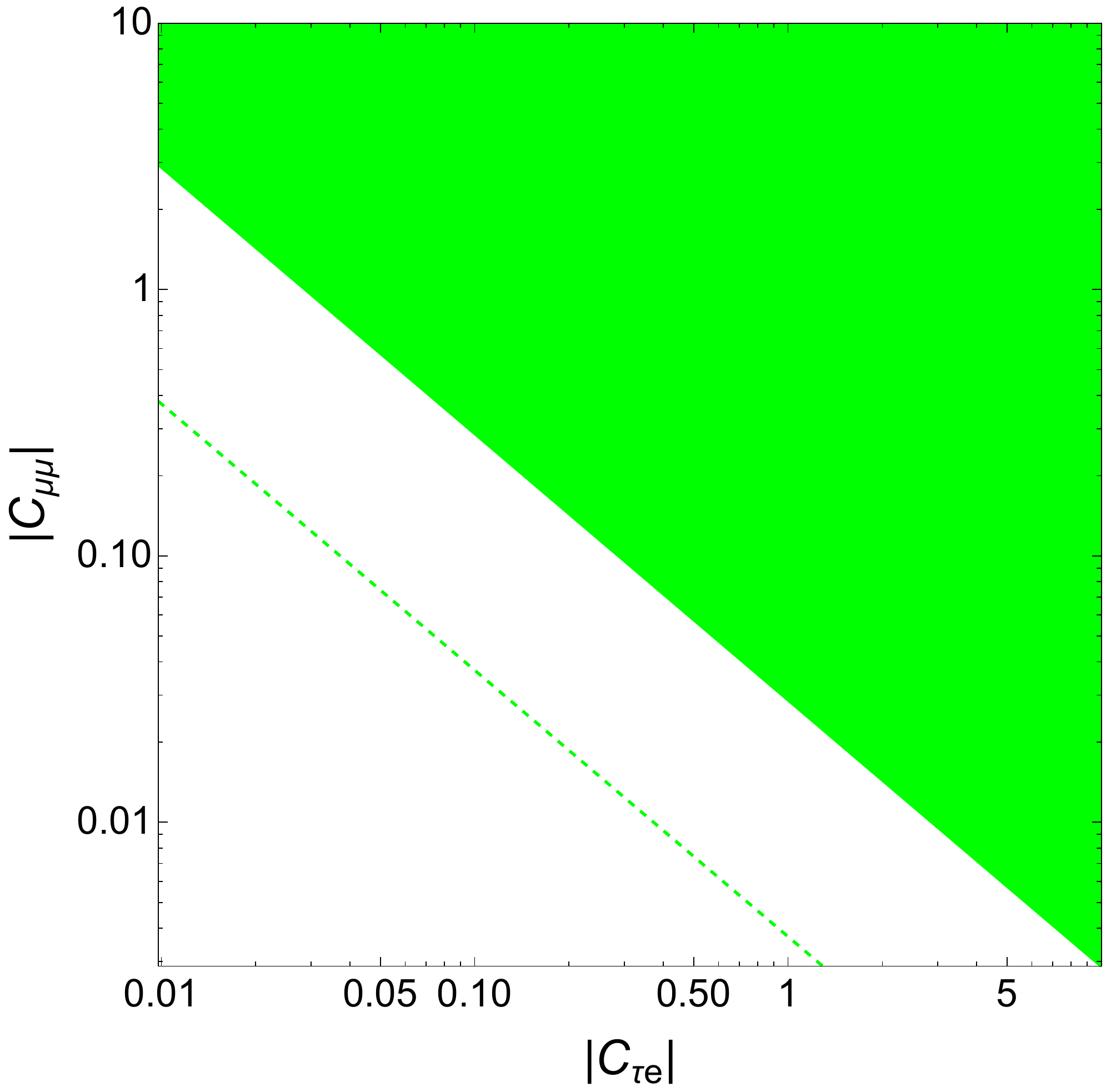}
}
\hspace{0mm}
\caption{\small The current experimental constraints and the future projections on the lepton FV through di-Higgs effective couplings $C_{ll'}$. The solid regions correspond to constraints, whereas the dashed lines represent future projections. In (a), the green color corresponds bounds/ projections from $\tau \rightarrow 3\mu$, whereas the blue corresponds to $\tau \rightarrow \mu\gamma$. In (b), the green corresponds to $\mu \rightarrow 3e$, and the blue to $\mu \rightarrow e\gamma$. In (c), the green corresponds to $\tau \rightarrow 3 e$, and the blue to $\tau \rightarrow e\gamma$. In (d), the green corresponds to $\tau^{-} \rightarrow e^{+}\mu^{-}\mu^{-}$, whereas the blue to $\tau^{-} \rightarrow \mu^{+}\mu^{-}e^{-}$, and the red to $\tau \rightarrow e\gamma$. In (e), the green arises from $\tau^{-}\rightarrow \mu^{+}e^{-}e^{-}$, the blue from $\tau^{-}\rightarrow \mu^{-}e^{+}e^{-}$, and the red from $\tau \rightarrow \mu\gamma$. In (f), the green arises from $\tau^{-}\rightarrow \mu^{+}\mu^{-}e^{-}$.}
\label{fig8}
\end{figure}

\begin{figure}[!t]
\centering
\subfloat[$|C_{ee}|$ Vs. $|C_{\tau\mu}|$]{
  \includegraphics[width=54mm]{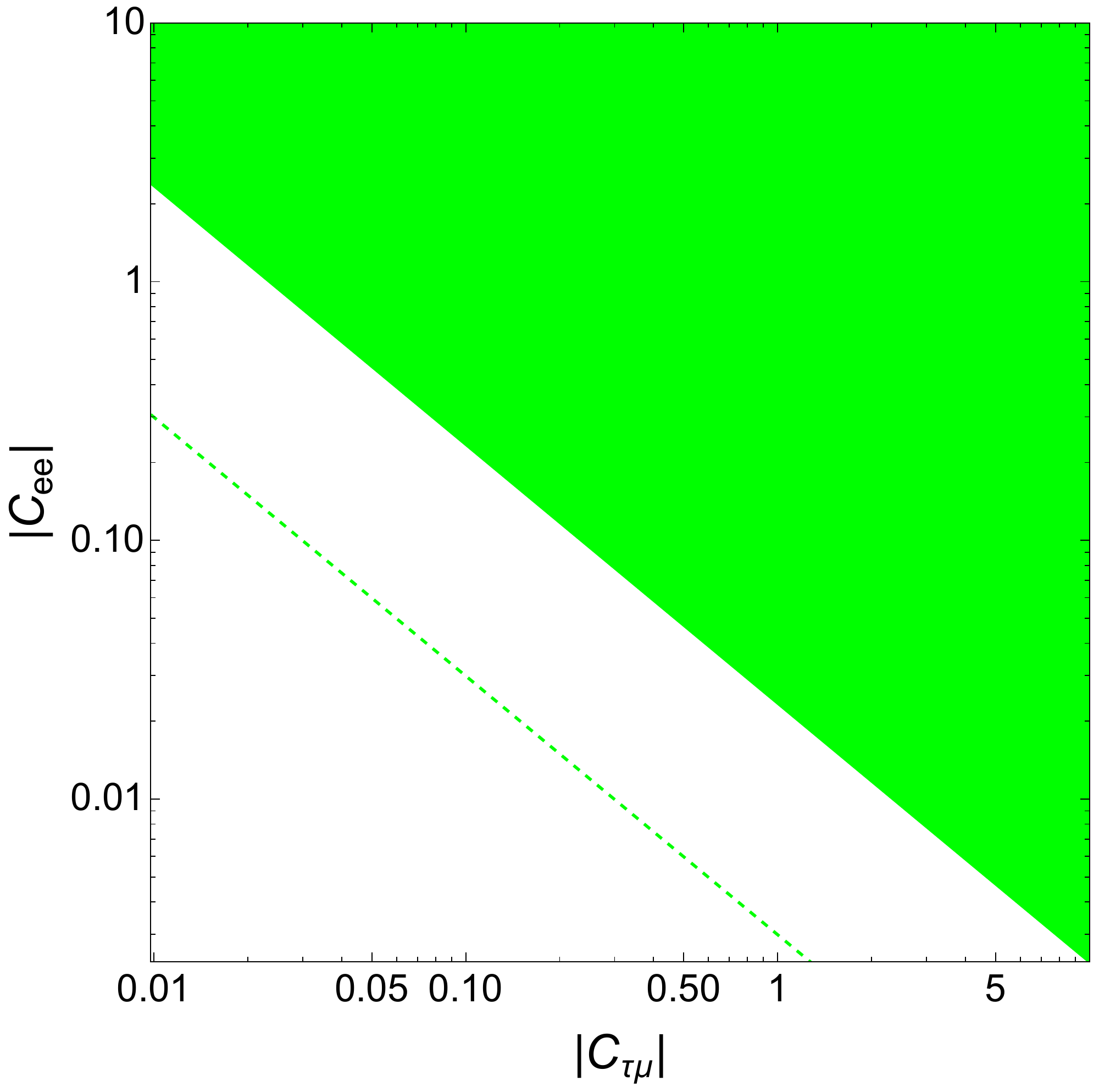}
}
\subfloat[$|C_{\tau e}|$ Vs. $|C_{\tau \mu}|$]{
  \includegraphics[width=54mm]{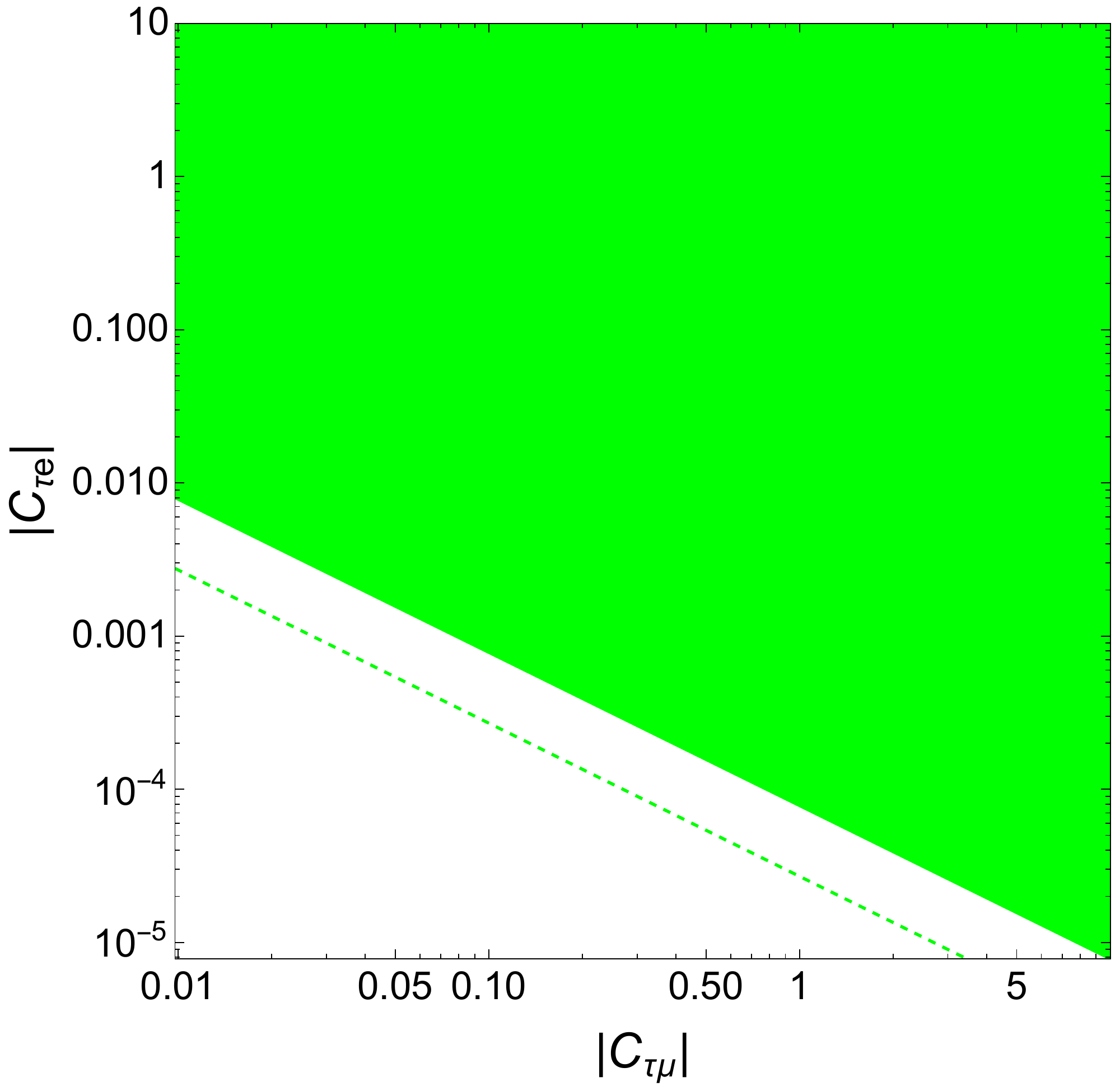}
}
\subfloat[$|C_{\mu\mu}|$ Vs. $|C_{\mu e}|$]{
  \includegraphics[width=54mm]{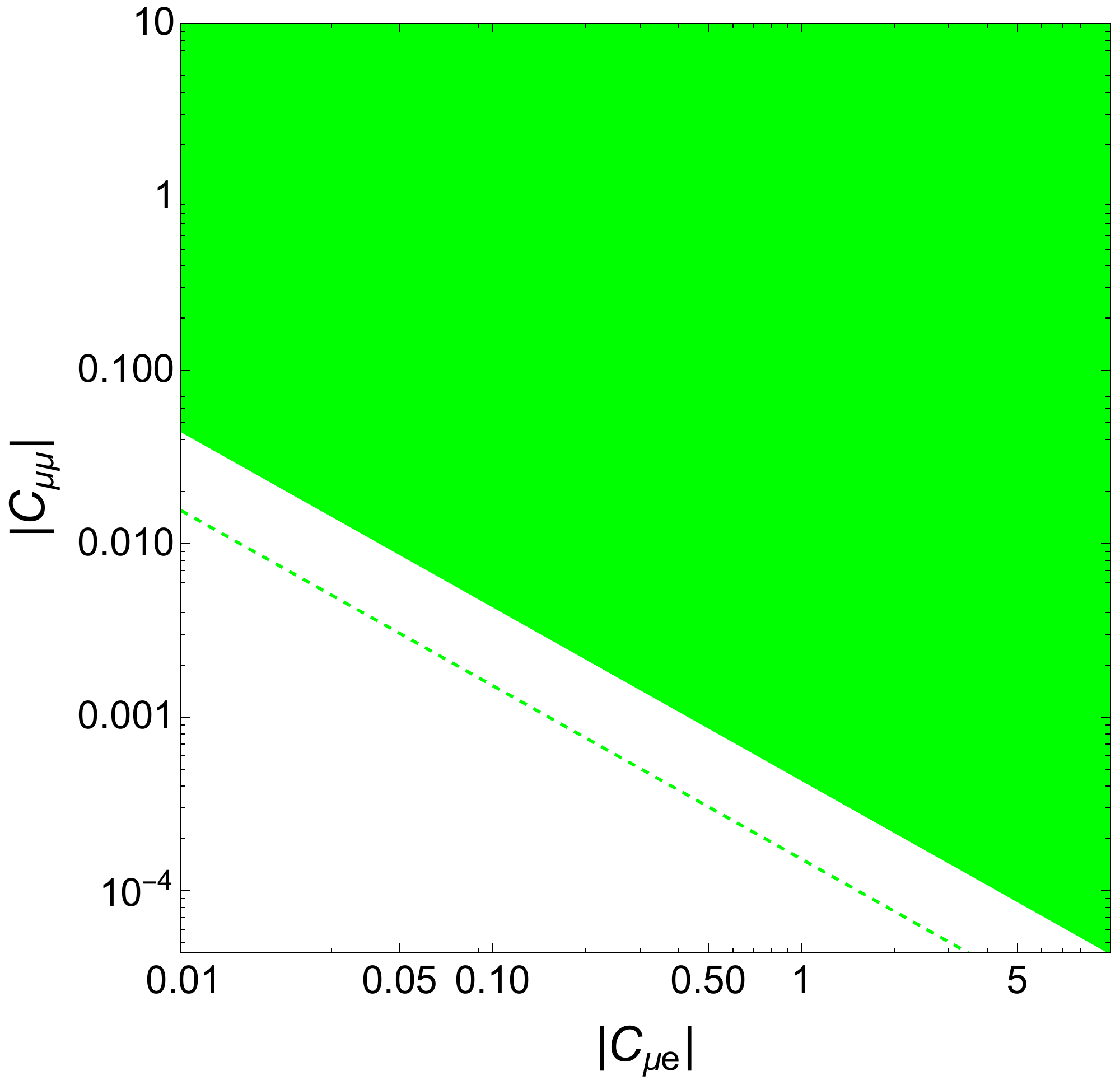}
}
\hspace{0mm}
\subfloat[$|C_{\tau\tau}|$ Vs. $|C_{\tau\mu}|$]{
  \includegraphics[width=54mm]{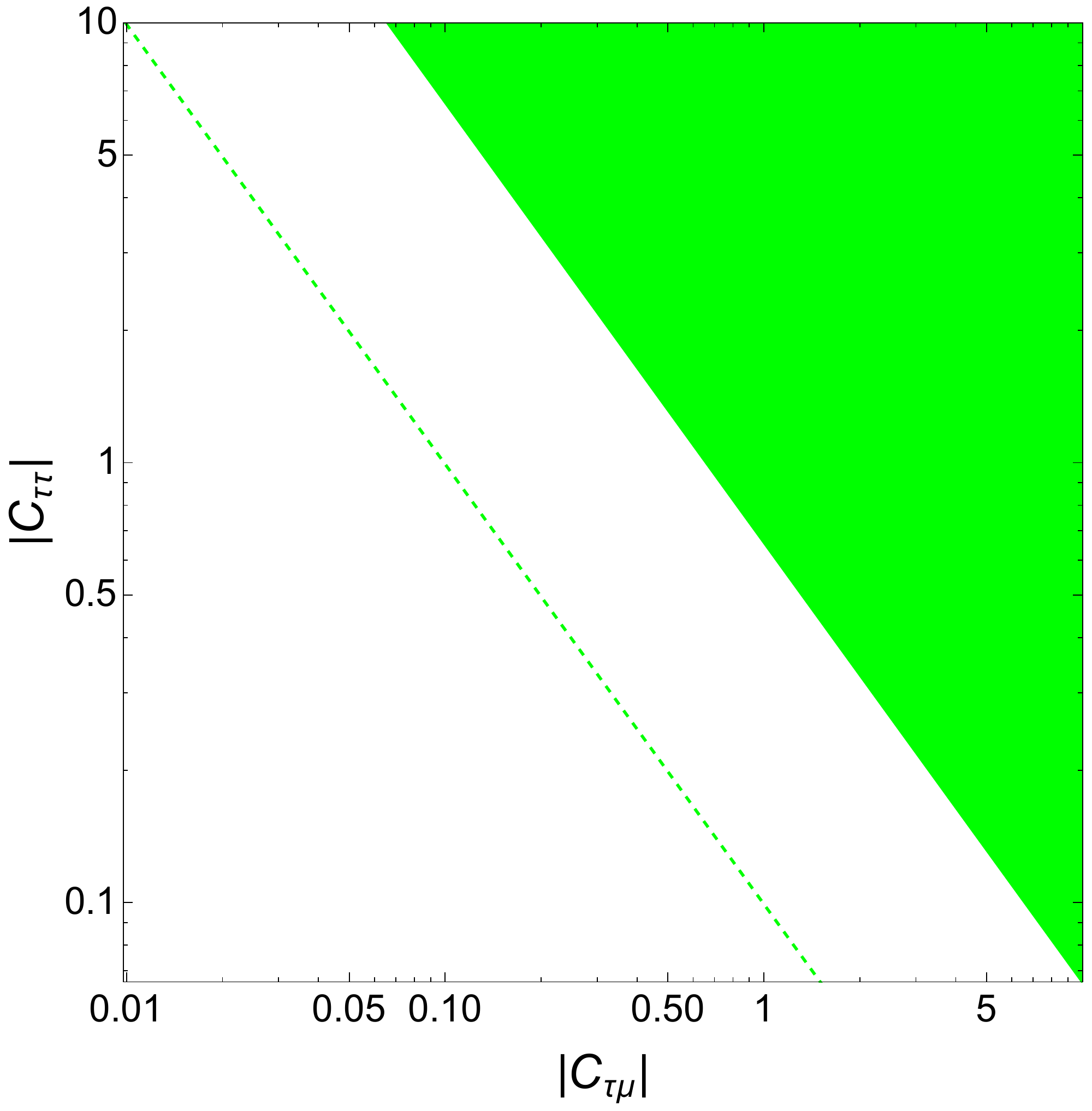}
}
\subfloat[$|C_{\tau\tau}|$ Vs. $|C_{\tau e}|$]{
  \includegraphics[width=54mm]{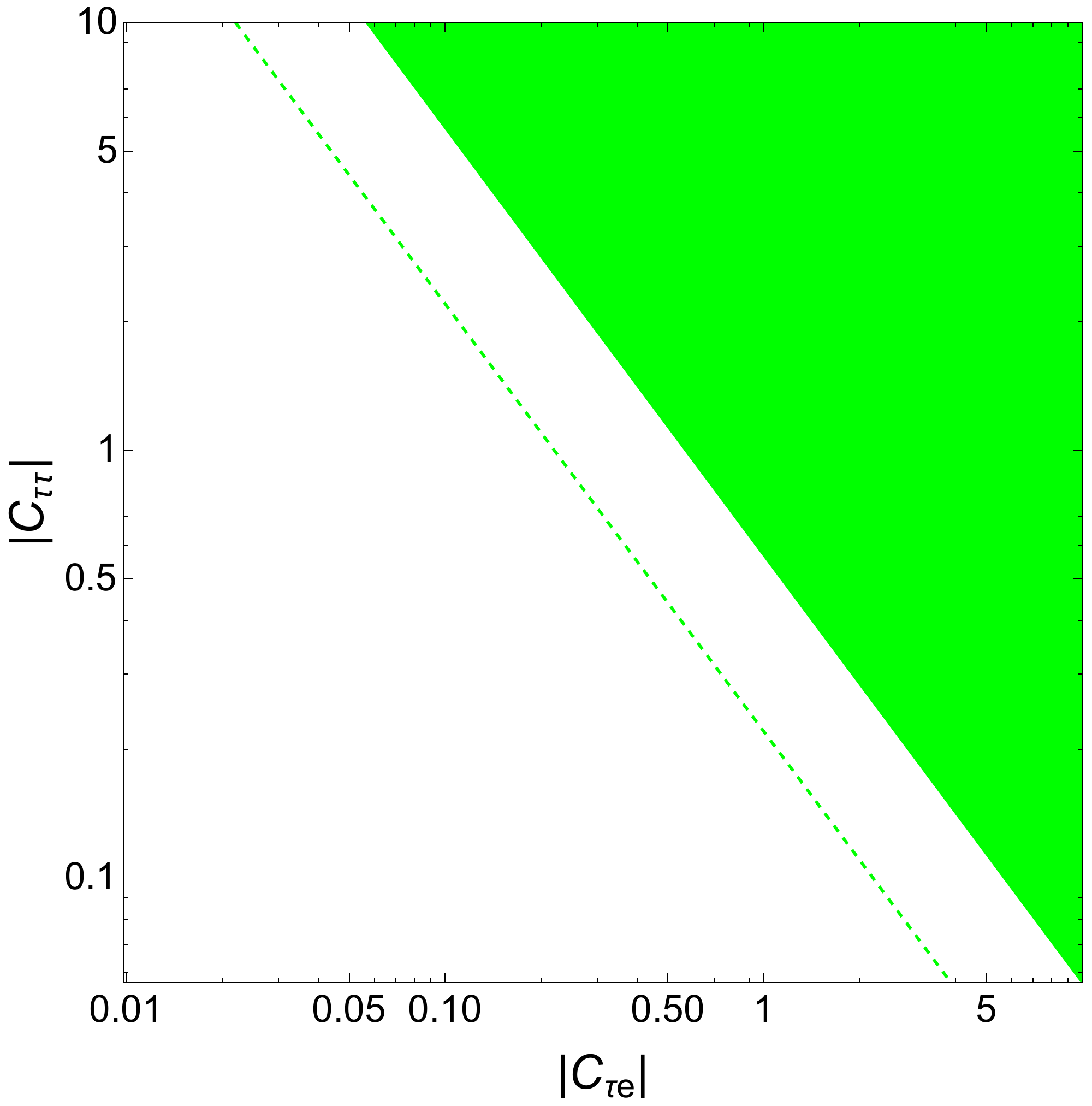}
}
\hspace{0mm}
\caption{\small (Cont.) The current experimental constraints and the future projections on the lepton FV through di-Higgs effective couplings $C_{ll'}$. The solid regions correspond to constraints, whereas the dashed lines represent future projections. In (a), the bounds and projections arise from $\tau^{-} \rightarrow \mu^{-}e^{+}e^{-}$, in (b) and (c) from $\mu \rightarrow e\gamma$, in (d) from $\tau \rightarrow \mu\gamma$ and in (e) from $\tau \rightarrow e\gamma$.}
\label{fig9}
\end{figure}

\subsection{Fine-tuning and lepton mass corrections}
Nonzero $C_{ll}$ can give rise to corrections to the masses of the leptons when the Higgs loop is closed. These corrections need to be suppressed in order to avoid the stringent bounds on the leptons' masses, which could lead to fine-tuning. We can easily estimate the level of fine-tuning associated with $C_{ll}$ as
\begin{equation}\label{eq:fine-tuning}
    \frac{\delta m_{l}}{m_{l}} \sim \frac{C_{ll}M_{h}^{2}}{32\pi^{2}v^{2}} \sim O(10^{-3}) \times C_{ll},
\end{equation}
which is negligible for the range of $C_{ll}$ required by FV constraints. Therefore, FV through $C_{ll}$ does not require any fine-tuning.

\section{Quark Sector}\label{sec:Quarks}
We now turn our attention to investigating the next-to-minimal FV couplings in the quark sector. We first discuss the constraints on the couplings $C_{ij}$ that arise from meson oscillations, then we investigate the bounds that can be extracted from $B$-physics searches. The constraints are summarized in Table \ref{table2}.
\begin{table}[!ht]
\centering
\vspace{1 mm}
\tabcolsep7pt\begin{tabular}{c c c c}
\hline
\hline
 \textbf{Channel} & \textbf{Couplings} &  \textbf{Bounds} & \textbf{$\Lambda$ (TeV)}\\
\hline
\hline
$D^{0}$ \text{Oscillations} & $|C_{uc}|$ & $< 7.73 \times 10^{-4}$ & 15.3 \\
$B^{0}_{d}$ \text{Oscillations} & $|C_{db}|$ & $< 1.73 \times 10^{-3}$ & 10.2\\
$B^{0}_{s}$ \text{Oscillations} & $|C_{sb}|$ & $< 1.50 \times 10^{-2}$ & 3.5\\
$K^{0}$ \text{Oscillations} & $|C_{sd}|$ & $< 1.20 \times 10^{-5}$ & 123 \\
\hline
$R_{K^{+}}$ & $|C_{\mu\mu}|/|C_{ee}|$ & $[0.93,1.01]$ & - \\
\hline
\hline
\end{tabular}
    \caption{\label{table2} \small $90\%$ CL bounds on the FV di-Higgs effective couplings in the quark sector and the corresponding lower limit on the scale of NP from matching to the SMEFT. The bounds are obtained from meson oscillations and $B$-physics searches.}
\label{table2}
\end{table}
\subsection{Constraints from meson oscillations}
Constraints on the couplings $C_{ij}$ can be obtained from meson oscillations, which include in particular $D^{0}-\overline{D}^{0}$, $B_{s,d}^{0}-\overline{B}_{s,d}^{0}$, and $K^{0}- \overline{K}^{0}$ oscillations. These oscillations can proceed through the di-Higgs couplings $C_{ij}$ via diagrams identical to the ones shown in Figure \ref{fig3}. The effective Hamiltonian of these diagrams can be written as \cite{UTfit:2007eik}
\begin{equation}\label{eq:Hamiltonian}
    \mathcal{H}_{\text{eff}} = C_{2,ij}(\overline{q}_{jR}^{\alpha}q_{iL}^{\alpha})(\overline{q}_{jR}^{\beta}q_{iL}^{\beta}) + C_{5,ij}(\overline{q}_{jR}^{\alpha}q_{iL}^{\beta})(\overline{q}_{jL}^{\beta}q_{iR}^{\alpha}),
\end{equation}
where $C_{2,ij}$ arises from integrating out the t-channel, whereas $C_{5,ij}$ arises from integrating out the s-channel in Figure \ref{fig3}. The detailed calculation of these loops is presented in Appendix \ref{appendix3}. In particular, the loop factor $V(P^{2})$ is given is Eq. (\ref{eq:fftoff2}), and in the non-relativistic limit where we can assume that $M_{h}^{2} \gg P^{2}$, $V(P^{2})$ is approximately given in Eq. (\ref{eq:fftoff4}). Thus, identifying the renormalization scale with the mass of the meson $m$, we can relate $C_{2,ij}$ and $C_{5,ij}$ defined in \cite{UTfit:2007eik} to the FV di-Higgs couplings $C_{ij}$ as follows
\begin{equation}\label{eq:C2C5}
    C_{2,ij} \simeq C_{5,ij} \simeq-\frac{i C_{ij}^{2}}{64\pi^{2}v^{2}}\log{\Big( \frac{M_{h}^{2}}{m^{2}}\Big)}.
\end{equation}

Using Eq. (\ref{eq:C2C5}) above, we can translate the bounds on $ C_{2,ij}$ and $C_{5,ij}$ presented in \cite{UTfit:2007eik} into bounds on $C_{ij}$.\footnote{The bounds presented in \cite{UTfit:2007eik} are @ $95\%$ CL. So, we rescale them to a $90\%$ C.L. as usual} $D^{0}-\overline{D}^{0}$ oscillations place constraints on the coupling $C_{uc}$. The stronger bound arises from $|C_{2,uc}|$ with an upper limit of $1.6 \times 10^{-13}$, which translates to the constraint $|C_{uc}| < 7.73 \times 10^{-4}$. $B_{d}^{0}-\overline{B}_{d}^{0}$, oscillations can set limits on the coupling $C_{db}$, where here, the stronger of the two bounds is $|C_{5,db}|< 6 \times 10^{-13}$, which translates into $|C_{db}| < 1.73 \times 10^{-3}$. On the other hand, $B_{s}^{0}-\overline{B}_{s}^{0}$, oscillations constrain the coupling $C_{sb}$, with $|C_{5,sb}| < 4.5 \times 10^{-11}$ being the more stringent bound, which leads to $|C_{sb}| < 1.5 \times 10^{-2}$. Finally, $K^{0}- \overline{K}^{0}$ oscillations place bounds on $C_{ds}$. These bounds however, only constrain the imaginary parts of $|C_{2,sb}|$ and $|C_{5,sb}|$. Specifically, the bounds read
\begin{align}
    \text{Im}(C_{2,ds}) & = [-5.1,9.3] \times 10^{-17}, \label{eq:ds_bound1}\\
    \text{Im}(C_{5,ds}) & = [-5.2,2.8] \times 10^{-17}.\label{eq:ds_bound2}
\end{align}

Given Eq. (\ref{eq:C2C5}) and our assumption that $C_{ij}$ are real, it's not hard to see that only the negative part of bounds in Eqs. (\ref{eq:ds_bound1}) and (\ref{eq:ds_bound2}) will be translated into a bound on $C_{ds}$. In addition, it's easy to see that $C_{2,ds}$ leads to a stronger bound, which translates to $|C_{ds}| < 1.2 \times 10^{-5}$.
\subsection{Bounds from $B-$physics}\label{sec:B_anomalies}
Historically, $B$-physics attracted a lot of attention because experimental searches revealed several discrepancies between their findings and the SM predictions. These flavor anomalies have stirred intensive research in $B$-physics (see \cite{London:2021lfn} for a recent review), however, recent experimental searches seem to eliminate most of these anomalies. In particular, the recent results from the LHCb \cite{LHCb:2022zom}, reveal that lepton universality ratios $R_{K^{+}}$ and $R_{K^{*}}$ are consistent with the SM model.

The lepton universality ratio $R_{K^{+}}$ is defined as 
\begin{equation}\label{eq:RK}
    R_{K^{+}} = \frac{\text{Br}(B^{+} \rightarrow K^{+}\mu^{+}\mu^{-})}{\text{Br}(B^{+} \rightarrow K^{+}e^{+}e^{-})}.
\end{equation}

\begin{figure}[!t]
\centering
  \includegraphics[width=80mm]{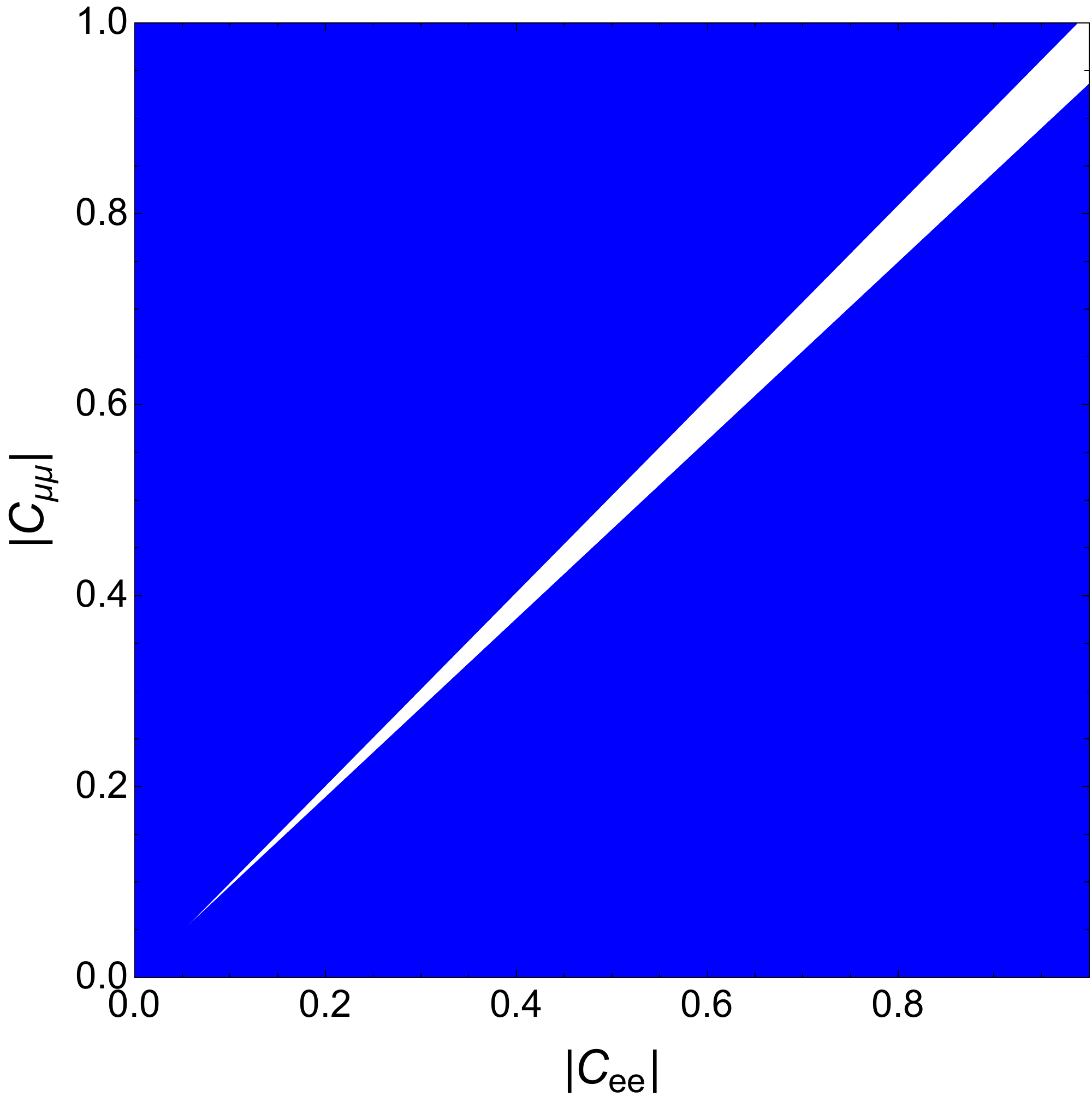}
\caption{\small The experimental bounds on the di-Higgs effective couplings extracted from $B$-physics. Specifically, the bound is extracted from the measurement of $R_{K^{+}}$.}
\label{fig10}
\end{figure}

At the quark level, the decay of the $B^{+}$ meson to $K^{+}l^{+}l^{-}$ with two leptons involves the decay $\overline{b} \rightarrow \overline{s}l^{+}l^{-}$, which can can proceed via di-Higgs couplings through a diagram similar to the ones in Figure \ref{fig3}. Given the results in Appendix \ref{appendix3}, it's easy to see that within our framework, $R_{K^{+}} = \frac{|C_{\mu\mu}|^{2}}{|C_{ee}|^{2}}$. The strongest bound on $R_{K^{+}}$ arises from the central $q^{2}$ region \cite{LHCb:2022zom}, with $R^{\text{Exp}}_{K^{+}} = 0.949^{+0.048}_{-0.047}$, which translates into the bound 
\begin{equation}
 \frac{|C_{\mu\mu}|}{|C_{ee}|} = [0.93,1.01] \hspace{1 cm} \text{@ 90\% C.L.}  
\end{equation}
The bound is shown in Figure \ref{fig10}.

\section{Matching to the SMEFT and the Scale of New Physics}\label{sec:SMEFT}
Finally in this section, we show how our framework matches to the SMEFT, then use the upper bounds on the FV Wilson coefficients to set lower limits on the corresponding scale of NP. Working in the Warsaw basis \cite{Grzadkowski:2010es}, there is only one class of operators at dimension-6 that contributes to the FV di-Higgs couplings, which has the form $\psi^{2}H^{3}$. There are 3 operator categories in $\psi^{2}H^{3}$
\begin{equation}\label{eq:SMEFT6}
    \mathcal{L}_{\psi^{2}H^{3}} = \frac{C^{l}_{ij}}{\Lambda^{2}}(H^{\dagger}H)(\overline{l}_{i}H e_{j}) + \frac{C^{u}_{ij}}{\Lambda^{2}}(H^{\dagger}H)(\overline{q}_{i}\Tilde{H} u_{j}) + \frac{C^{d}_{ij}}{\Lambda^{2}}(H^{\dagger}H)(\overline{q}_{i}H d_{j}) + h.c.,
\end{equation}
which should be matched to the operators in eq. \ref{FV_Cij}. The matching is identical for all of the operators and is fairly straightforward: We simply plug the Higgs doublet in eq. \ref{eq:SMEFT6}, then we match the $h^{2}$ term to eq. \ref{FV_Cij}. setting $C_{ij}^{\text{SMEFT}} = 1$, we find
\begin{equation}\label{eq:matchingC}
    C_{ij} = \frac{3 v^{2}}{\Lambda^{2}}.
\end{equation}

Eq. \ref{eq:matchingC} can be used to set a lower limit on the scale of NP $\Lambda$ from the upper bounds on $C_{ij}$. We present these bounds in Tables \ref{table1} and \ref{table2}. In the lepton sector, we can see that lower bounds ranges between $\sim 1-10$ TeV. On the other hand, the stronger bounds in the quark sector lead to much higher scales $\Lambda$, ranging from a few TeV to $\sim 123$ TeV. 

\section{A Possible UV Completion}\label{sec:UVcomplletion}
Here we present a possible UV completion for the FV di-Higgs couplings. In general, we need a UV completion where the leading contribution to FV (from $hff$ couplings) is suppressed, whereas the NLO contribution (from $hhff$ couplings) is allowed to be sizable. To this avail, we adopt a model similar to the one used in \cite{Dermisek:2021mhi}. We limit ourselves to the lepton sector. Extension to the quark sector should be straightforward.

In this UV completion, we extend the SM by 2 vector-like $SU(2)$ doublet fermions $L_{L,R}$, and 2 vector-like $SU(2)$ singlet fermions $E_{L,R}$. Here $E$ and $L$ have the same quantum numbers as the SM leptons. Such models with vector-like fermions have been used extensively in the literature for a variety of purposes, such as solving the $g-2$ anomaly (see for instance \cite{Dermisek:2021mhi,Kannike:2011ng, Dermisek:2013gta, Poh:2017tfo, Dermisek:2021ajd, Arkani-Hamed:2021xlp}). In addition, we extend the SM by a Higgs-like scalar doublet that develops a VEV $v_{\phi}$.
\begin{equation}\label{eq:Phi_doublet}
\Phi = \begin{pmatrix}
0\\
v_{\phi} + \phi
\end{pmatrix}.
\end{equation} 
 
\begin{figure}[!t]
\centering
\includegraphics[width=120mm]{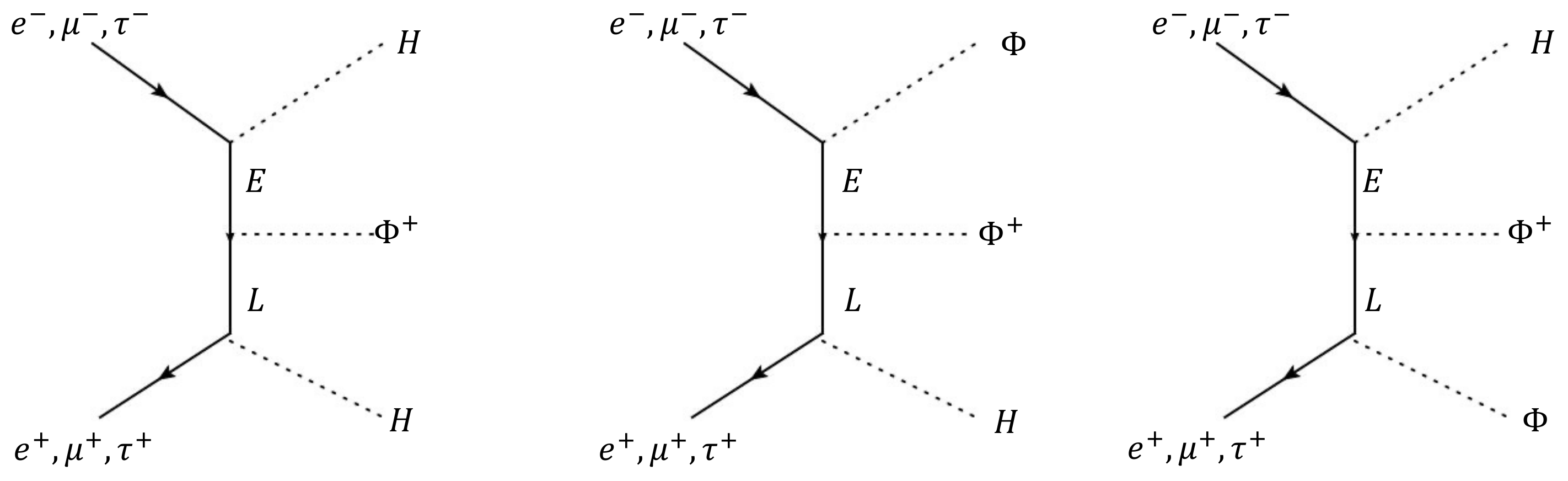}
\caption{\small Feynman diagrams of the UV completion.}
\label{fig11}
\end{figure}

The most general Lagrangian reads
\begin{equation}\label{eq:UV_Lagrangian}
\begin{split}
\mathcal{L}_{\text{UV}} = & -M_{L}\overline{L}_{L}L_{R} -M_{E}\overline{E}_{L}E_{R} - y_{i}\overline{l}_{L_{i}}H e_{R_{i}} - Y_{i} \overline{l}_{L_{i}} \Phi e_{R_{i}}-\lambda_{E_{i}}\overline{l}_{L_{i}}H E_{R} -\lambda_{L_{i}}\overline{L}_{L}H e_{R_{i}} \\
& - \lambda\overline{L}_{L}HE_{R} - \lambda^{\prime} \overline{E}_{L} H^{\dagger}L_{R} -\kappa_{E_{i}}\overline{l}_{L_{i}}\Phi E_{R} - \kappa_{L_{i}}\overline{L}_{L}\Phi e_{R_{i}} - \kappa \overline{L}_{L}\Phi E_{R} - \overline{\kappa}\overline{E}_{L}\Phi^{\dagger} L_{R} + \text{h.c.},
\end{split}
\end{equation}
where $l_{L}$ is a L-handed lepton doublet and $e_{R}$ is a R-handed lepton singlet, and $ i = e, \mu, \tau$. In our calculation, we decouple $\Phi$ by assuming that it is too heavy and we set $\lambda^{\prime} \rightarrow 0$.  Assuming that $M_{L}, M_{E} \gg m_{\text{EW}}$, we can integrate out $L$ and $E$ (see Figure \ref{fig11}) to arrive at the effective Lagrangian
\begin{equation}\label{eq:UV_effective}
	\mathcal{L}_{\text{EFT}} \supset - y_{i}\overline{l}_{L_{i}}H e_{R_{i}} - Y_{i} \overline{l}_{L_{i}}\Phi e_{R_{i}} -\frac{\lambda_{L_{i}}\lambda_{E_{j}}}{\overline{M}^{2}}\overline{l}_{L_{i}}H\Phi^{\dagger}H e_{R_{j}} - \Bigg( \frac{\kappa_{E_{i}}\lambda_{L_{j}}+ \lambda_{E_{i}}\kappa_{L_{j}}}{\overline{M}^{2}} \Bigg)\overline{l}_{L_{i}}\Phi\Phi^{\dagger}H e_{R_{j}}+ h.c.,
\end{equation}
where we have defined $\overline{\kappa}/M_{L}M_{E} \equiv 1/\overline{M}^{2}$. We can plug the Higgs and $\Phi$ doublets in eq. (\ref{eq:UV_effective}) and match it to the effective Lagrangian, but before doing so, notice that the lepton mass matrix will receive diagonal corrections when $i=j$, in addition to off-diagonal corrections when $i \neq j$. Upon diagonalizing the lepton mass matrix, these corrections can be suppressed by tuning them against $Y_{i}$, which are free parameters, so that these corrections remain within the acceptable experimental limits. Now, matching eq. (\ref{eq:UV_effective}) to the FV Lagrangian in eq. (\ref{eq:lepton_FV}), and requiring that the tree-level FV contribution be suppressed, we obtain the following matching conditions
\begin{eqnarray}\label{eq:matching}
\lambda_{L_{i}} & \simeq & -\frac{v_{\phi}(\lambda_{E_{i}}\kappa_{L_{j}} + \kappa_{E_{i}} \lambda_{L_{j}})}{\sqrt{2}v \lambda_{E_{j}}},\nonumber \\ 
C_{ij} & = & \frac{\sqrt{2}v v_{\phi}\lambda_{L_{i}}\lambda_{E_{j}}}{\overline{M}^{2}},
\end{eqnarray}
which can easily be accommodated for a suitable choice of parameters. For example, in order to obtain $|C_{\mu e}| \sim 1.59 \times 10^{-3}$ as required by the bounds in Table \ref{table1}, we can use $v_{\phi} = 5$ TeV, $\kappa_{L_{e}} = 0.1$, $\lambda_{E_{\mu}} = 0.2$, $\kappa_{E_{\mu}}=0.1$, $\lambda_{L_{e}} = -0.15$ and $\lambda_{E_{e}} = 0.3$, which would require $\lambda_{L_{\mu}} = -0.24$ to accommodate the first matching condition in eq. (\ref{eq:matching}), and using $\overline{M} = 10$ TeV in the second matching condition yields the required value for  $|C_{\mu e}|$.

Notice however, that such a scenario does require some fine-tuning to suppress the LO contribution, which is inevitable sans some hidden symmetry that forces such a cancellation. Nonetheless, this is to be expected in scenarios where the LO contribution is suppressed compared to the NLO, and even exists in FV in the Higgs sector at tree-level \cite{Harnik:2012pb}.
\section{Conclusions}\label{sec:conclusions}
In this paper, we employed a completely model-independent bottom-up EFT to investigate FV in the quark and the lepton interactions with the Higgs. In this approach we dubbed the WSD, we did not resort to any power expansion, and instead listed the most general FV interactions. This approach is a generalization of the one introduced in \cite{Chang:2019vez, Abu-Ajamieh:2020yqi, Abu-Ajamieh:2021egq,Abu-Ajamieh:2022ppp, Abu-Ajamieh:2021vnh} to the FV case.

Unlike previous studies on FV in the Higgs sector which focused on FV Yukawa couplings $Y_{ij} \neq \sqrt{2}m_{i}\delta_{ij}/v$. In this paper, we focused on the next-to-minimal FV that arises from the di-Higgs effective couplings of the form $h^{2}\overline{f}f$ and assumed that the Yukawa couplings are equal to the SM predictions. To the best our knowledge, this is the first time constraints are set on these types of couplings.

In the lepton sector, we investigated the bounds on the FV di-Higgs couplings that arise form $l \rightarrow l_{1}l_{2}l_{3}$ decays, $l_{i} \rightarrow l_{k}\gamma$ decays, muonium oscillations, the $g-2$ anomaly, LEP searches, $\mu \rightarrow e$ conversion in nuclei, leptonic FV Higgs decays, and from both flavor-conserving and FV $Z$ decays. We have set upper limits on both individual effective couplings and products of the various couplings and found that these bounds in general vary between $\sim O(1)$ down to $\sim O(10^{-5})$. In addition, we utilized the projections of some future experiments, such as Belle II, the Mu2e experiment and the HL-LHC in order to find future projections on some of these couplings and found that these bounds can be improved by roughly a factor ranging between a few and two orders of magnitude. The bounds and future projections are summarized in Table \ref{table1} and Figures \ref{fig8} and \ref{fig9}. On the other hand, bounds on the FV di-Higgs couplings in the quark sector were obtained from meson oscillations and from $B$-physics searches, and these range between $\sim O(10^{-2})$ down to $\sim O(10^{-5})$. These bounds are summarized in Table \ref{table2} and Figure \ref{fig10}. 

We have shown how our approach can be mapped to the SMEFT and have shown the scale of NP that corresponds to the upper limits on the FV Wilson coefficients. We saw that the scale of NP ranges between $\sim 1- 10$ TeV in the lepton sector, and between a few TeV up to $\sim 123$ TeV in the quark sector. We believe that measuring the di-Higgs effective couplings, whether flavor-conserving or FV, is of particular importance and should receive adequate attention in the LHC searches and other low-energy experiments. The proposed muon collider would be an interesting laboratory where these couplings can be probed. 

\section*{Acknowledgments}
 The work of F.A. is supported by the C.V. Raman fellowship from CHEP at IISc. S.K.V. thanks SERB Grant CRG/2021/007170 "Tiny Effects from Heavy New Physics"  and Matrics grant MTR/2022/000255, "Theoretical Aspects of Certain Physics Beyond Standard Models" from the Department of Science and Technology, Government of India.

\appendix
\section{The decay width of $l \rightarrow l_{1}l_{2}l_{3}$}\label{appendix1}
The matrix element of the decay shown in Figure \ref{fig1} is given by
\begin{equation}\label{eq:mat1}
    \mathcal{M}_{123} = -i \frac{C_{ll_{1}}C_{l_{2}l_{3}}}{2v^{2}}\overline{u}(q_{1}) \int \frac{d^{4}k}{(2\pi)^{4}}\frac{1}{(k-p+q_{1})^{2}-M^{2}}\frac{1}{k^{2}-M^{2}}
    u(p)\overline{u}(q_{3})v(q_{2}),
\end{equation}
where $M$ is the mass of the particle in the loop. The loop is logarithmically divergent and needs regularization. We use DR to perform the momentum integral
\begin{equation}\label{eq:mat2}
   \mathcal{M}_{123} =  \frac{C_{ll_{1}}C_{l_{2}l_{3}}}{32 \pi^{2}v^{2}}\overline{u}(q_{1}) u(p)\overline{u}(q_{3})v(q_{2}) \Gamma(2-\frac{d}{2}) \Big( \frac{\mu^{2}}{M^{2}}\Big)^{2-\frac{d}{2}} \int_{0}^{1}dx \Big[ 1+x(x-1)(m^{2}+m_{1}^{2}-2p.q_{1})/M^{2}\Big]^{\frac{d}{2}-2},
\end{equation}
where $m, m_{1}$ are the masses of $l, l_{1}$ respectively and $\mu$ is the renormalization scale. Before we perform the integral over the Feynman parameter, we notice that in the limit $M \gg m, m_{1}$ applicable in our case, we can drop the masses $m$ and $m_{1}$ from the integral. In addition, in the rest frame of the decaying particle, $p.q_{1} = m E_{1}$, with $E_{1}$ being the energy of $l_{1}$. Given that the upper limit of $E_{1}$ is $m/2$, we can drop that term as well. Therefore, in the limit $M \gg m, m_{1}$ the integral in Eq. (\ref{eq:mat2}) becomes trivial. Setting $d = 4-2\epsilon$ and using the $\overline{\text{MS}}$ scheme, the regularized matrix element reads
\begin{equation}\label{eq:mat2}
\mathcal{M}_{123} = -\frac{C_{ll_{1}}C_{l_{2}l_{3}}}{32 \pi^{2}v^{2}} \log{\Big( \frac{M^{2}}{m^{2}}\Big)} \overline{u}(q_{1}) u(p)\overline{u},(q_{3})v(q_{2}),
\end{equation}
where have set $\mu^{2} = m^{2}$. Before we calculate the decay width, we point out that depending on the decay, there could be either one or two Feynman diagrams. For example, $\tau \rightarrow 3\mu$ obviously involves only one Feynman diagram, i.e., $l = \tau$, and $l_{1},l_{2},l_{3} = \mu$. On the other hand, a process like $\tau^{-} \rightarrow \mu^{+}\mu^{-}e^{-}$ involves two diagram: the first with $l_{1} = e^{-}$, $l_{2} = \mu^{+}$, and $l_{3} = \mu^{-}$, and the second with the $l_{1}$ and $l_{3}$ interchanged. The matrix elements of the two diagrams should be added together, with the appropriate Fermi-Dirac statistics taken into consideration. Here we show the decay width of processes with only one Feynman diagram. Generalizing to processes with two Feynman diagrams is straightforward. 

Since the decays we are interested in are $\tau \rightarrow 3\mu$, $\tau \rightarrow 3e$, $\tau \rightarrow \mu\mu e$, $\tau \rightarrow \mu ee$ and $\mu \rightarrow 3e$, in all cases we have $m \gg m_{1},m_{2},m_{3}$. Thus, we can treat the final states as massless. This simplifies the phase space integral greatly and the final result reads
\begin{equation}\label{lto3lDecayApp}
    \Gamma(l \rightarrow l_{1}l_{2}l_{3}) = \frac{m^{5}}{v^{4}}\Bigg[\frac{C_{l l_{1}}C_{l_{2}l_{3}}}{512\pi^{3}\sqrt{6\pi}} \log{\Big( \frac{M_{h}^{2}}{m_{l}^{2}}\Big)} \Bigg]^{2}.
\end{equation}

\section{Calculating the 2-loop diagram of $l_{i} \rightarrow l_{k}\gamma$}\label{appendix2}
Here we show the general calculation of the 2-loop diagram in Figure \ref{fig2}. This diagram is the leading contribution to the decays $\tau \rightarrow \mu \gamma$, $\tau \rightarrow e \gamma$ and $\mu \rightarrow e \gamma$. Notice that in each case, the inner particle $j$ could be either $\tau$, $\mu$, or $e$, which leads to different structures of the matrix element with the corresponding effective couplings $C_{ij}$ and $C_{jk}$. We can write the matrix element as 
\begin{equation}\label{eq:ltolgamma1}
 \mathcal{M}_{ijk} = \frac{e C_{ij}C_{jk}}{2v^{2}} \overline{u}(p-q,m_{k})I_{ijk}u(p,m_{i})\epsilon_{\mu}^{*}(q),
\end{equation}
where the two-loop momentum integral is given by
\begin{equation}\label{eq:ltolgamma2}
 I_{i,j,k}  = \int \frac{d^{4}k_{1}}{(2\pi)^{4}} \frac{(\slashed{k}_{1}-\slashed{q}+m_{j})\gamma^{\mu}(\slashed{k}_{1}+m_{j})}{[(k_{1}-q)^{2}-m_{j}^{2}][k_{1}^{2}-m_{j}^{2}]}\int \frac{d^{4}k_{2}}{(2\pi)^{4}}\frac{1}{[(p-k_{1}-k_{2})^{2}-M^{2}][k_{2}^{2}-M^{2}]}.
\end{equation}

We can perform the integral over $k_{2}$ first, then combine the results with the remaining integral over $k_{1}$, and finally perform the momentum integral over $k_{1}$. Using DR, we find the following general form of the matrix element
\begin{align}\label{eq:ltolgamma3}
 \mathcal{M}_{ijk} & = \frac{e C_{ij}C_{jk}}{(4\pi)^{4} v^{2}} \Gamma(4-d)(-1)^{5-\frac{d}{2}}\Big( \frac{4\pi \mu^{2}}{M_{h}^{2}}\Big)^{4-d}\overline{u}(p-q,m_{k})u(p,m_{i})(p\cdot \epsilon^{*}) \\ \nonumber
 & \times \int_{0}^{1}dx \int_{0}^{1}dy \int_{0}^{1}dz z^{1-\frac{d}{2}}(1-z)\frac{\Big[ (a+b-1)(a\hspace{0.5mm} m_{i}+m_{j})-b(a\hspace{0.5mm} m_{k}+m_{j})\Big]}{\alpha^{\frac{d}{2}}\beta^{4-d}},
\end{align}
where $M_{h}$ is the mass of the Higgs, $p^{\mu}$ the momentum of the initial state lepton, and the functions $\alpha$, $\beta$, $a$ and $b$ are given by
\begin{align}\label{eq:ltolgamma4}
    a & = \frac{xz(x-1)}{\alpha},\\
    b & = \frac{y(z-1)}{\alpha}, \\
    \alpha & = (x^{2}-x+1)z-1,\\
    \beta & = -z + \frac{x z (x-1)(y-1)(z-1)}{\alpha} \frac{m_{i}^{2}}{M_{h}^{2}} + (z-1)\frac{m_{j}^{2}}{M_{h}^{2}} - \frac{x y z (x-1)(z-1)}{\alpha} \frac{m_{k}^{2}}{M_{h}^{2}}.
\end{align}

The integrals in Eq. (\ref{eq:ltolgamma3}) are badly divergent and care is needed to regularize them. In addition, it is not possible to evaluate them exactly for any general particles $i, j$ and $k$. Thus, we need to approximate them by assuming $M_{h} \gg m_{\tau} \gg m_{\mu} \gg m_{e}$, and only keep the lepton with the largest mass in each decay. Notice that in Eq. (B7), although $M \gg m_{i,j,k}$, we need to keep the term with the largest lepton mass to keep the integral IR finite. Therefore, evaluating Eq. (\ref{eq:ltolgamma3}) will depend on what the particles $i$, $j$ and $k$ are. In order to set upper limits on the FV couplings $C_{ij}$, we treat each case separately. For example, for the process $\tau \rightarrow \mu$, we could have $j = {\tau, \mu, e}$ running in the loop. This furnished 9 distinct processes in total to consider. Here we show a sample calculation, then quote the results for the rest of the process.

Consider the process $\tau \rightarrow e \gamma$ with $\mu$ in the loop. We denote the corresponding matrix element by $\mathcal{M}_{\tau \mu e}$, with $m_{i} = m_{\tau}$, $m_{j} = m_{\mu}$ and $m_{k} = m_{e}$. Dropping $m_{\mu}, m_{e}$, the integral in Eq. (\ref{eq:ltolgamma3}) simplifies to 
\begin{equation}\label{eq:ltolgamma5}
 \mathcal{M}_{\tau\mu e} \simeq m_{\tau} \int_{0}^{1}dx \int_{0}^{1}dy \int_{0}^{1}dz z^{1-\frac{d}{2}}(1-z)\frac{a(a+b-1)}{\alpha^{\frac{d}{2}}\beta^{4-d}},
\end{equation}
with 
\begin{equation}\label{eq:ltolgamma6}
\beta \simeq -z + \frac{x z (x-1)(y-1)(z-1)}{\alpha} \frac{m_{\tau}^{2}}{M^{2}}.
\end{equation}

The integral in Eq. (\ref{eq:ltolgamma5}) is still divergent. So, in order to regularize it, we use the method described in \cite{Peskin:1995ev}. First, we define the function
\begin{equation}\label{eq:ltolgamma6}
f(z) \equiv (1-z)\frac{a(a+b-1)}{\alpha^{\frac{d}{2}}\beta^{4-d}}.
\end{equation}
then isolate the divergence by splitting the integral over $z$ as follows
\begin{equation}\label{eq:ltolgamma7}
f(z)= \int_{0}^{1}dx \int_{0}^{1}dy \Bigg[ \int_{0}^{1}dz z^{1-\frac{d}{2}} f(0) +  \int_{0}^{1}dz z^{1-\frac{d}{2}} \Big( f(z) - f(0) \Big)\Bigg] = -\frac{1}{6}.
\end{equation}

Plugging Eq. (\ref{eq:ltolgamma7}) in Eq. (\ref{eq:ltolgamma3}), then setting $d=4-2\epsilon$ and using the $\overline{\text{MS}}$ scheme, we arrive at final answer
\begin{equation}\label{eq:ltolgamma8}
\mathcal{M}_{\tau \mu e} \simeq -\frac{e C_{\tau\mu}C_{\mu e}m_{\tau}}{6(4\pi)^{4}v^{2}}\overline{u}_{e}(p-q)u_{\tau}(p) (p\cdot\epsilon^{*})\log{\Big( \frac{M_{h}^{2}}{m_{\mu}^{2}} \Big)},
\end{equation}
where we have set the renormalization scale $\mu^{2} = m_{\mu}^{2}$ in the logarithm.

\section{$f_{i}\overline{f}_{j} \rightarrow f_{k}\overline{f}_{l}$ scattering}\label{appendix3}
Here we show how to calculate the matrix element of the scattering $f_{i}\overline{f}_{j} \rightarrow f_{k}\overline{f}_{l}$, which will be used to find the bounds from LEP, muonium-antimuonium oscillations and meson oscillation. At 1-loop, the scattering proceeds through the s- and t-channels as in Figure (\ref{fig2}). The matrix element is given by
\begin{align}\label{eq:fftoff1}
    i\mathcal{M} & = i\mathcal{M}_{s} + i\mathcal{M}_{t}, \nonumber \\
    &  =\frac{C_{ij}C_{kl}}{4v^{2}} \overline{u}_{k}(k_{1})v_{l}(k_{2}) \overline{v}_{j}(p_{2})u_{i}(p_{1})V(P_{s}^{2}) - \frac{C_{ik}C_{jl}}{4v^{2}} \overline{u}_{k}(k_{1})u_{i}(p_{1})\overline{v}_{j}(p_{2})v_{l}(k_{2}) V(P_{t}^{2}),
\end{align}
where $P_{s} = p_{1}+p_{2}$, $P_{t} = p_{1}-k_{1}$, and $p_{1,2}$ ($k_{1,2}$) are the initial (final) momenta. The loop integral is given by
\begin{equation}\label{eq:fftoff2}
    V(P^{2}) = \int \frac{d^{4}k}{(2\pi)^{4}}\frac{1}{(k+P)^{2}-M^{2}}\frac{1}{k^{2}-M^{2}}.
\end{equation}

The integral in Eq. (\ref{eq:fftoff2}) is logarithmically divergent and needs regularization. The suitable choice of regularization will depend on the type of process at hand. In high energy scattering like in LEP, using a UV cutoff is more appropriate. Evaluating the integral using a UV cutoff $\Lambda$, the final result can be approximated by
\begin{equation}\label{eq:fftoff3}
    V(P^{2}) \simeq \frac{i}{16\pi^{2}}\Bigg( 1 + \log{\Big( \frac{\Lambda^{2}}{M^{2}}\Big)} +\sqrt{1-\frac{4M^{2}}{P^{2}}} \log{ \Bigg[ \frac{\sqrt{1-4M^{2}/P^{2}}-1}{\sqrt{1-4M^{2}/P^{2}}+1}\Bigg] }\Bigg).
\end{equation}

On the other hand, in the non-relativistic limit suitable for $M-\overline{M}$ and meson oscillation, it is more suitable to evaluate the integral using DR. In the $\overline{\text{MS}}$ scheme, the integral evaluates to
\begin{equation}\label{eq:fftoff4}
    V(P^{2})  \simeq \frac{i}{16\pi^{2}}\log{\Big(\frac{\mu^{2}}{M_{h}^{2}} \Big)},
\end{equation}
where $\mu$ is the renormalization scale. Notice that in the non-relativistic limit $M^{2}\gg P^{2}$, Eq. (\ref{eq:fftoff4}) can be obtained from Eq. (\ref{eq:fftoff3}) by taking the limit $P^{2} \rightarrow 0$ and then setting $\Lambda^{2} = e \mu^{2} $.

\section{Detailed calculation of $\mu \rightarrow e$ conversion in nuclei}\label{appendix4}
The most general effective Lagrangian can be expressed as \cite{Kitano:2002mt}
\begin{multline}\label{eq:m2eLag}
    \mathcal{L}_{\text{eff}} = c_{L} \frac{e}{8\pi^{2}}m_{\mu}(\overline{e}\sigma^{\mu\nu}P_{L}\mu)F_{\mu\nu} -\frac{1}{2}\sum_{q}\Big[ g^{q}_{LS}(\overline{e}P_{R}\mu)(\overline{q}q) +g^{q}_{LP}(\overline{e}P_{R}\mu)(\overline{q}\gamma_{5}q)  \\ +g^{q}_{LV}(\overline{e}\gamma^{\mu}P_{L}\mu)(\overline{q}\gamma_{\mu}q)
    +g^{q}_{LA}(\overline{e}\gamma^{\mu}P_{L}\mu)(\overline{q}\gamma_{\mu}\gamma_{5}q) +\frac{1}{2}g^{q}_{LT}(\overline{e}\sigma^{\mu\nu}P_{R}\mu)(\overline{q}\sigma_{\mu\nu}q)\Big] + (L \leftrightarrow R),
\end{multline}
where the sum is over all quarks. Here, the first term expresses the contributions arising from the magnetic dipole operators as in the bottom diagram of Figure \ref{fig2}. On the other hand, the terms inside the square brackets refer to the scalar, pseudoscalar, vector, pseudo-vector and tensor contributions, respectively. As shown in Figure \ref{fig5}, only the scalar and tensor contributions are non-vanishing. Furthermore, the tensor contribution is expected to be small, and the bounds are not expected to compete with those from $l_{i} \rightarrow l_{k}\gamma$, therefore we neglect it as well.

The scalar contribution $g^{q}_{LS}$ and $g^{q}_{RS}$, are shown in the left diagram of Figure \ref{fig5}. They can be calculated by integrating out the loop in the non-relativistic limit and at vanishing momentum transfer, yielding 
\begin{equation}\label{eq:mu2eScalr}
    g^{q}_{LS} = g^{q}_{RS} \equiv g^{q}_{S} = \frac{3\sqrt{2}C_{\mu e}Y_{q}^{2}m_{N}}{64\pi^{2}v M_{h}^{2}}. 
\end{equation}
where $Y_{q}$ is the quark Yukawa coupling and $m_{N}$ is the mass of the nucleon. The $\mu \rightarrow e$ conversion rate receives contributions from protons and neutrons and can be expressed as \cite{Kitano:2002mt}
\begin{equation}\label{eq:mu2eConvRate}
    \Gamma(\mu \rightarrow e) = |\Tilde{g}^{(p)}_{S} S^{(p)} + \Tilde{g}^{(n)}_{S} S^{(n)}|^{2},
\end{equation}
where
\begin{equation}\label{eq:g_tilde}
    \Tilde{g}^{(p)}_{S} = \sum_{q}g^{q}_{S}\frac{m_{p}}{m_{q}} f^{(q,p)}, \hspace{2cm} \Tilde{g}^{(n)}_{S} = \sum_{q}g^{q}_{S}\frac{m_{n}}{m_{q}} f^{(q,n)},
\end{equation}
where the nucleon matrix elements $f^{(q,N)} \equiv \bra{N}m_{q}\overline{q}q\ket{N}/m_{N}$. These nucleon matrix elements were calculated in \cite{Ellis:2008hf} but using an older value for the nucleon sigma term $\Sigma_{\pi N} =64$ MeV. Using the updated value of $59.6$ MeV \cite{Gupta:2021ahb}\footnote{In \cite{Harnik:2012pb}, the nucleon matrix elements were calculated using the then latest value of $\Sigma_{\pi N} =55$ MeV, however, there is an error in their equation A19. In particular, $f^{(u,n)} = 0.018 \neq f^{(d,p)}$, and $f^{(d,n)} = 0.043 \neq f^{(u,p)}$. All other values were correctly calculated for $\Sigma_{\pi N} =55$ MeV.}, the nucleon matrix elements for the light quarks are given by
\begin{align}
    f^{(u,p)} \simeq  0.022, \hspace{1cm} f^{(d,p)} \simeq  0.038, \hspace{1cm} f^{(s,p)} \simeq  0.342, \label{eq:p_nucleon_element}\\
    f^{(u,n)} \simeq  0.018, \hspace{1cm} f^{(d,n)} \simeq  0.049, \hspace{1cm} f^{(s,n)} \simeq  0.342, \label{eq:n_nucleon_element}
\end{align} 
whereas the contribution for the heavy quarks is obtained from
\begin{equation}\label{eq:eq:nucleon_heavy}
    f^{(c,N)} = f^{(b,N)} = f^{(t,N)} = \frac{2}{27} \Big(1 - \sum_{q = u, d,s} f^{(q,N)} \Big) \simeq 0.044,
\end{equation}
for both the neutron and proton. The coefficients $S^{(p)}$, $S^{(n)}$ are the overlap integrals of the electron, muon and nuclear wavefunctions for the proton and neutron respectively. They are tabulated for a variety of target materials in \cite{Kitano:2002mt}. According to  \cite{SINDRUMII:2006dvw}, gold provides the strongest bound on the conversion rate
\begin{equation}\label{eq:mu2e_bound}
    \text{Br}^{\text{Au}}(\mu \rightarrow e) = \Bigg[ \frac{\Gamma (\mu \rightarrow e)}{\Gamma^{\mu}_{\text{Capture}}}\Bigg]_{\text{Au}} < 7 \times 10^{-13} \hspace{5mm} @ \hspace{1mm} 90\% \hspace{2mm} \text{C.L.},
\end{equation}
and we find from \cite{Kitano:2002mt} that $\Gamma^{\text{Au}}_{\text{Capture}} = 13.07 \times 10^{6} \hspace{1mm} \text{s}^{-1}$. In addition, the overlap coefficients for gold are given by $S^{(p)} = 0.0614$ and $S^{(n)} = 0.0918$ in units of $ m_{\mu}^{5/2}$. On the other hand, the Mu2e experiment is projected to improve the measurement of the conversion rate by roughly 3 orders of magnitude through utilizing aluminum as its stopping material. More specifically, the projected bound of the Mu2e experiment is given by \cite{Kargiantoulakis:2019rjm}
\begin{equation}\label{eq:mu2e_projection}
    \text{Br}^{\text{Al}}(\mu \rightarrow e) = \Bigg[ \frac{\Gamma (\mu \rightarrow e)}{\Gamma^{\mu}_{\text{Capture}}}\Bigg]_{\text{Al}} <  10^{-16} \hspace{5mm} @ \hspace{1mm} 90\% \hspace{2mm} \text{C.L.},
\end{equation}
and we have $\Gamma^{\text{Al}}_{\text{Capture}} = 0.7054 \times 10^{6} \hspace{1mm} \text{s}^{-1}$, and the overlap coefficients for aluminum are given by $S^{(p)} = 0.0155$ and $S^{(n)} = 0.0167$ in units of $ m_{\mu}^{5/2}$.

\end{document}